\documentclass[a4paper,fleqn]{cas-dc}
\usepackage[square,sort,comma,numbers]{natbib}
\setcitestyle{square}

\usepackage{graphicx} 
\usepackage{fancyvrb}
\usepackage{forest}

\usepackage{tikz}
\usetikzlibrary{trees,positioning,shapes,shadows,arrows}
\usepackage{todonotes}

\usepackage{amsmath}
\usepackage{amssymb}
\usepackage{hyperref}
\usepackage{mathrsfs}
\usepackage{circledsteps}
\usepackage{threeparttable}
\usepackage{xspace}
\usepackage{enumitem}

\usepackage{soul}
\usepackage{xcolor}
\usepackage{amssymb}
\usepackage{pifont}

\usepackage{amsthm}
\usepackage{makecell}

\theoremstyle{definition}
\newtheorem{definition}{Definition}[section]

\newcommand{\prover}{$\mathcal{P}$\xspace}
\newcommand{\proverprogram}{$T_{\mathcal{P}}$\xspace}
\newcommand{\verifier}{$\mathcal{V}$\xspace}
\newcommand{\verifierprime}{$\mathcal{V}'$\xspace}
\newcommand{\directquote}[1]{\emph{``#1''}\xspace}

\newcommand{\partialreq}[1]{#1\P\xspace}

\newcommand{\hwinstrumentation}{$\blacktriangle$\xspace}
\newcommand{\uncertain}[1]{\emph{#1}$\dag$\xspace}
\newcommand{\cmark}{\ding{51}}%
\newcommand{\xmark}{\ding{55}}%

\usepackage{amsmath} 

\usepackage[T1]{fontenc}

\newcounter{serialnumber}
\pgfkeys{/csteps/inner color=black}
\pgfkeys{/csteps/fill color=white}
\newcommand{\npapers}{31\xspace}

\begin{document}
\let\WriteBookmarks\relax
\def\floatpagepagefraction{1}
\def\textpagefraction{.001}

\shorttitle{Control-Flow Attestation: Concepts, Solutions, and Open Challenges}

\shortauthors{Z.\ Sha et~al.}

\title [mode = title]{Control-Flow Attestation: Concepts, Solutions, and Open Challenges}                    


%
\author[1]{Zhanyu Sha}[type=editor,
                        auid=000,bioid=1,
                        style=chinese,
                        orcid=0009-0002-2706-7136]

\cormark[1]


\ead{zhanyu.sha.2022@rhul.ac.uk}


\affiliation[1]{organization={Information Security Group},
    addressline={Royal Holloway, University of London}, 
    city={Egham},
     state={Surrey},
    country={United Kingdom}}

\affiliation[2]{organization={School of Computing},
    addressline={Newcastle University}, 
    city={Newcastle-upon-Tyne},
     state={Tyne and Wear},
    country={United Kingdom}}

\author[2]{Carlton Shepherd}[orcid=0000-0002-7366-9034]
\ead{carlton.shepherd@ncl.ac.uk}
%
\author[1]{Amir Rafi}[]
\ead{amir.rafi.2019@live.rhul.ac.uk}
\author[1]{Konstantinos Markantonakis}[]
\ead{k.markantonakis@rhul.ac.uk}



\cortext[cor1]{Corresponding author}

\tnotetext[1]{Carlton Shepherd has received funding from the UK EPSRC `Chameleon' project (EP/Y030168/1).}


\begin{abstract}
Control-flow attestation unifies the worlds of control-flow integrity and platform attestation by measuring and reporting a target's run-time behaviour to a verifier. Trust assurances in the target are provided by testing whether its execution follows an authorised control-flow path. The problem has been explored in various settings, such as assessing the trustworthiness of cloud platforms, cyber-physical systems, and Internet of Things devices. Despite a significant number of proposals being made in recent years, the area remains fragmented, with different adversarial behaviours, verification paradigms, and deployment challenges being addressed.  In this paper, we present the first survey of control-flow attestation, examining the core ideas and solutions in state-of-the-art schemes. In total, we survey over 30 papers published between 2016--2024, consolidate and compare their key features, and pose several challenges and recommendations for future research in the area.
\end{abstract}


\maketitle

\section{Introduction} \label{Introduction}

Today's computing platforms are susceptible to attacks which manipulate software behaviour at run time. \emph{Control-flow attacks}, such as return-oriented programming (ROP), have become commonplace that execute unintended program code paths in order to perform unauthorised functions using existing code fragments. These fragments are combined in sequence in a way that evades conventional defences, such as code signing and other static integrity-checking methods. The challenge of run-time threats has underpinned the area of \emph{control-flow integrity} (CFI), which aims to ensure that programs adhere to authorised paths of execution~\cite{RN170}. This can be achieved, for example, by verifying the legitimacy of the destination address of every control-flow transfer (e.g.\ jump and call instructions). Other mechanisms have also been developed, such as \emph{code pointer integrity} (CPI)~\cite{RN177}, which segregates pointers from other data in memory; and \emph{data execution prevention} (DEP), which prevents code from being executed from regions of memory that are not explicitly marked as executable.

In this paper, we concern ourselves with the problem of \emph{control-flow attestation} (CFA). In general, CFA schemes aim to bridge the problems of CFI and platform attestation by building assurances that a target device has not been compromised at run-time. This contrasts to conventional remote attestation that relies on measuring static components, such as bootloaders and other software binaries~\cite{abera2016things}. CFA overcomes the limitations of static attestation, which cannot reliably detect run-time attacks \emph{after} any measurements have been taken.  A typical CFA scheme involves a proving device, \prover, and a challenger or verifier, \verifier. Given an input challenge from \verifier, \prover creates a measurement that is an aggregation of different steps in the resulting program execution. These `steps' may correspond to the different nodes that were traversed within the target's \emph{control flow graph} (CFG), where \verifier knows the set of legitimate paths before \prover's deployment.

In recent years, a large number of CFA schemes have been proposed in the literature, focusing on different measurement approaches, roots of trust, adversarial models, and hardware and software assumptions. For instance, different hardware security technologies have been suggested for protecting \prover's measurement and reporting procedures~\cite{cflat,gonzalez2024lightfat,oat,RN169}; various static and dynamic methods have been proposed for program instrumentation~\cite{cflat,RN156,RN158,RN139,RN145}; and numerous control-flow transfer types have been monitored~\cite{RN157,RN169,scarr,RN139,yadav2023whole}. Despite receiving significant attention from the research community, no survey exists that consolidates and reflects on the state of the art. In this paper, we address this gap through a comprehensive review of existing CFA proposals and make the following contributions:

\begin{itemize}
\item We present the first systemisation of knowledge paper on control-flow attestation, covering 31 CFA proposals over approximately the last decade since CFA was first introduced in the literature.
\item We consolidate the knowledge areas explored in state-of-the-art CFA schemes, and describe their application in existing proposals.
\item Using a common set of criteria, we compare existing CFA schemes with respect to their methods, trust assumptions, target control-flow transfers, and adversarial models in order to elucidate their features, advantages, and shortcomings.
\end{itemize}

\subsection{Related Work}
 To the best of our knowledge, no extensive survey currently exists summarising CFA schemes; however, several existing papers survey remote attestation (RA) more generally in different contexts, which we summarise in Table~\ref{tab:related-surveys}.
 
 In 2016, Steiner \& Lupu~\cite{steiner2016attestation} presented one of the first surveys covering traditional remote attestation approaches applicable to wireless sensor networks. Arias et al.~\cite{arias2018device} (2018) provided a short review of device attestation methods for IoT devices, touching principally on software, hardware, and swarm-based attestation in the static measurement model. In 2020, Sfyrakis \& Gro{\ss}~\cite{sfyrakis2020survey} reviewed different hardware approaches for attesting devices in network infrastructures, such as those used in infrastructure-as-a-service solutions. In 2021, Johnson et al.~\cite{RN120} proposed a taxonomy of RA schemes for embedded systems, categorising proposals according to their root of trust, evidence type, evidence gathering, packaging and verification, and scalability. Ankerg{\aa}rd et al.~\cite{ankergaard2021state} examined software-only remote attestation schemes without relying on hardware-based roots of trust for IoT devices. Kuang et al.~\cite{RN127} (2022) survey remote attestation schemes for IoT devices in general, detailing attestation proposals targeting different attacks (e.g.\ static and run-time software attacks, and physical attacks). Ambrosin et al.~\cite{RN121} review solutions in the area of collective remote attestation, where multiple devices are attested as a whole rather than individually. Recently, Ammar et al.~\cite{ammar2024sok} (2024) provided a summary of runtime integrity mechanisms, focusing on CFI and a selection of CFA papers, discussing their commonalities and distinctions, such as architecture, scalability, and attack vectors. However, their work does not offer a comprehensive review of state-of-the-art CFA schemes, which is the focus of our analysis.

\begin{table}
\caption{An overview of related survey papers.}
\centering
\resizebox{\columnwidth}{!}{%
\begin{tabular}{r|l|l|c}
\toprule
\textbf{Work} & \textbf{Year} & \textbf{Description} & \makecell{\textbf{\# CFA}\\\textbf{Papers}}\\ \midrule
Steiner \& Lupu~\cite{steiner2016attestation} & 2016 & RA methods for wireless sensor networks. & 1 \\
Arias et al.~\cite{arias2018device} & 2018 & A brief review of device attestation methods. & 2 \\
Sfyrakis \& Gro{\ss}~\cite{sfyrakis2020survey} & 2020 & Hardware methods for attesting network infrastructures.  & 0 \\
Johnson et al.~\cite{RN120}   &   2021   &  A review of RA schemes for embedded systems.  & 6 \\
Ankerg{\aa}rd et al.~\cite{ankergaard2021state}  &    2021           &   State-of-the-art software-based RA for IoT devices.  & 2                  \\
Kuang et al.~\cite{RN127}    &    2022           &   A general review of RA schemes for IoT devices.  & 8                 \\
Ambrosin et al.~\cite{RN121}    &    2022           &  Collective RA schemes for large IoT networks. & 5  \\
Ammar et al.~\cite{ammar2024sok} & 2024 & A review of runtime integrity mechanisms & 18
\\\midrule
\emph{This Paper}     & \emph{2024} & \emph{A comprehensive analysis of control-flow attestation.} & 31 \\ \bottomrule
\end{tabular}
}
\label{tab:related-surveys}
\end{table}

 In several surveys~\cite{arias2018device,RN120,RN121,ammar2024sok}, control-flow attestation is described briefly, focusing on early control-flow attestation schemes (e.g.~\cite{cflat,RN156}).  Survey papers on control-flow integrity more generally have also been published on verification methods~\cite{RN131} and hardware-based CFI systems~\cite{RN171}. To the best of our knowledge, we are not aware of a dedicated treatment of CFA proposals that have emerged in recent years. We directly address this gap in this paper.

\subsection{Scope}

To scope this work, we focus on the problem of attesting a target platform by collecting evidence derived from control-flow events. We introduce definitions based on the principles and definitions established in~\cite{coker2011principles}.

\begin{definition}[Verifier and Target]
    A \emph{verifier} is an entity, such as a computer on a network, that makes a decision about another party. A \emph{target} or \emph{prover} is a party, such as an untrusted device, about which a verifier must make such a decision.
\end{definition}

The verifier checks whether the target system state meets certain security requirements by analyzing evidence provided by the target. This evidence serves as proof in a process known as \emph{attestation}, which helps the verifier determine if the target is free from unauthorised changes, malware, or other violations that may compromise security.

 \begin{definition}[Attestation and Verification]
\emph{Attestation} is the activity of making a claim to a verifier about the properties of a target by supplying evidence which supports that claim. \emph{Verification} refers to the verifier's decision-making process based on the information it receives during the attestation process from the prover.
\label{def:verification}
\end{definition}

\begin{definition}[Measurement]
    A proving device collects evidence about a target to be attested through direct and local observation using its \emph{measurement} process. The verifier uses its verification process to evaluate this evidence.
    \label{def:measurement}
\end{definition}

Substantial work has been conducted in generating evidence using \emph{static} measurements, such as cryptographic hashes of firmware or software binaries~\cite{sfyrakis2020survey,RN120,RN127,steiner2016attestation,ankergaard2021state,arias2018device,RN121}. While static attestation provides a snapshot of the target's state at a given point in time, it lacks the ability to detect run-time attacks, such as control-flow hijacking (see~\S\ref{sec:cfa-attacks}), which may occur after the target is under execution and been deemed trustworthy.

To address these limitations, the area of control-flow attestation has emerged, where evidence is collected at run-time during the target's execution from its control-flow behaviour. In this work, we conducted an extensive literature search using multiple databases, such as IEEE Xplore, ACM Digital Library, and Google Scholar. We used keywords and phrases closely associated with the topic area, including ``run-time'', ``control-flow'', ``dynamic'' and ``hybrid'' with the terms ``attestation'', ``trust'', and ``trusted computing''. We also incorporated forward and backward citation tracking to identify additional relevant works, particularly using related survey papers in Table~\ref{tab:related-surveys}. Papers were selected if they satisfied the following criteria: \begin{itemize}
    \item  The work proposed a new method of attestation aimed at establishing trust in a target entity, involving a prover and a verifier.
    \item  The proposal incorporated control-flow information produced at run-time by the target (prover) within an attestation response during the target's execution.  
\end{itemize}

As such, we are agnostic to the particular instruction set architecture, manufacturer or device under consideration.  Proposals that focus on control-flow integrity, but not attestation, are considered useful supplementary literature (see~\cite{burow2017control}). However, such proposals are ultimately considered outside the  scope of this work. Similarly, static attestation methods, and those that rely on a combination of static attestation and run-time integrity enforcement, e.g.~\cite{kucab2023hardware}, are also excluded.


\subsection{Paper Structure}

We begin with preliminary information in \S\ref{Background}, covering the principles behind control-flow integrity, the basics of control-flow graphs, and control-flow and non-control-data attacks addressed by CFA schemes. After this, we begin introducing the core ideas behind CFA in \S\ref{sec:introducing-cfa}, such as definitions and the knowledge areas of state-of-the-art proposals. We then systematically discuss each area of this taxonomy, discussing prover-verifier paradigms in \S\ref{sec:prover-verifier-paradigms}; trust anchors in \S\ref{sec:trust-anchors}; and methods for instrumenting and measuring proving programs in \S\ref{sec:instrumentation} and \S\ref{Measurement}. We consolidate existing CFA proposals according to these features in \S\ref{Summary} and present a comparison table to aid future research in the area. Lastly, we present several open research problems and recommendations in \S\ref{sec:open-problems} before concluding the paper in \S\ref{sec:conc}.

\section{Fundamentals of Control-Flow Integrity} \label{Background}

The goal of control-flow integrity is to ensure that a given program follows only the intended code paths. Put otherwise, it aims to ensure that program execution adheres to a set of policies. A typical method of enforcing CFI is through the use of a \emph{control-flow graph} (CFG) derived from the program's source code, binary, or an intermediate representation. A CFG represents all potential, and authorised, execution paths that a program can follow during its runtime. A CFG for a given program, $P$, can be represented as a directed graph, $G = (V,E)$, where:

\begin{itemize}
    \item $V$ is a set of vertices, where each vertex, $v_i \in V$, represents an independent sequence of instructions.
    \item $E$ is a set of edges where each edge is defined as a tuple, $(v_i, v_j) \in E$, representing a possible control flow transfer between two vertices. Transfers can be the result of function calls, returns, or (un-)conditional jumps in $P$.
\end{itemize}

Vertices of a CFG are referred to as \emph{basic blocks} (BBLs), which are considered an atomic sequence of program instructions with a single entry and exit point. The entry point is the first instruction of a BBL, while the exit refers to the final instruction. A control-flow transfer instruction, such as a branch or jump, is typically considered the last instruction of a BBL. Other instructions within a BBL do not alter the control flow. As such, the execution of a BBL is linear, with each instruction following the previous one until the final instruction. A control-flow transfer will have a source corresponding to the exit point of a BBL, and a destination that points to the entry point of a BBL. Given a control transfer instruction at runtime from a point ($p$) in a block ($v_i$) to a target address ($t$), the transfer is allowed iff $\exists(v_i, v_j) \in E$ in the CFG such that $t$ is the starting address of block $v_j$.

When the CFI security property~\cite{RN170} holds for a given program, $P$, we say that $P$ adheres to an authorised walk, $w = \{v_0, v_1, \dots v_n\}$, for a connected sequence of vertices with starting and final vertices, $v_0$ and $v_n$ respectively.\footnote{Existing work often uses the term `path' instead of `walk', even though this can be considered a misnomer in graph theory. A \emph{path} is a trail in which all vertices (and thus all edges) are distinct, whereas a \emph{walk} is more general: it is a finite or infinite sequence of edges which joins a sequence of vertices.} For a given program, the implementation of control-flow integrity is achieved generally in two phases:

\begin{enumerate}

    \item \emph{CFG construction}: The target program's CFG is computed using static or dynamic analysis or a combination thereof. Static analysis uses the program code using its binary, source code, or an intermediate form (cf.\ LLVM's intermediate representation~\cite{muntean2019analyzing}). This process is used to identify all possible control flow transfers before its execution. In contrast, dynamic analysis builds the CFG through giving inputs and logging program execution.    
    \item \emph{CFI enforcement}: This is performed at runtime and involves a reference monitor that checks and enforces the program's adherence to the CFG derived in the previous step. A common CFI enforcement method is through program instrumentation~\cite{RN170}. Here, a checkpoint $C(p,t)$ is inserted at a control transfer instruction at point $p$ in BBL, with a target, $t$. Here, $p$ and $t$ represent a logical CFG edge between one BBL $v_i$ and a target BBL, $v_j$. Before executing any control transfer instruction, the program passes control to the reference monitor that verifies whether $C(p,t)$ is associated with $(v_i, v_j) \in E$. That is, the monitor ensures that the transfer complies with the CFG. If the check fails, the monitor may halt the program's execution, throw an exception, or take another enforcement action. 
\end{enumerate}

In practice, control-flow transfers are instigated by instructions that alter the normal sequential execution flow, directing it from the current instruction to a different, non-consecutive address in the program. A control-flow transfer has two main attributes: a \emph{source} address and a \emph{destination} address. Transfers are divided into forward and backward ones, which redirect execution to a new location and return execution to the prior location respectively. On X86 platforms, the \texttt{call} and \texttt{ret} instructions are two examples of such instructions.  Forward control-flow transfers may further be categorised into direct and indirect calls and jumps. Branching and jump instructions may be executed according to some condition, e.g.\ \texttt{jz} (jumps if the zero flag, ZF, is set) and \texttt{jnle} (jump if not less or equal).

\subsection{Control-Flow Attacks}
\label{sec:cfa-attacks}

Control-flow integrity and attestation schemes aim to address the problem of control-flow attacks that manipulate control-flow transfers in order to perform unauthorised actions. In this section, we describe the common attack classes, and label them for later comparison in \S\ref{Summary}.

\subsubsection{Control Data Attacks} \label{Control data attacks}

The primary class are attacks which directly modify control data that influences the processor's program counter (PC) during execution by modifying return addresses and function pointers~\cite{RN125}.  A fundamental attack vector, which we refer to as code injection, involves inserting malicious code into executable memory and directing the processor to execute that code.

\begin{enumerate}[start=1,leftmargin=1cm,label={(A.\arabic*)}]
        \item \textbf{Code injection attacks:} \label{lab:code injection attacks} The attacker directly inserts malicious code into the target program's memory, e.g.\ through improper input validation and sanitisation, which is subsequently executed.
 \end{enumerate} 
    
Stack-based buffer overflows are a canonical example, which involves overwriting function return addresses and other variables on the stack to redirect execution to code, such as a malicious payload (e.g.\ shellcode). Heap-based attacks are another approach, involving overflowing buffers on the heap and affecting adjacent heap metadata and objects (e.g.\ function pointers) that redirect control flow.

More modern attacks involve the concept of \emph{code reuse}, which rely on unauthorised sequences of existing authorised code fragments in order to realise malicious functionality.

\begin{enumerate}[start=2,leftmargin=1cm,label={(A.\arabic*)}]
        \item \textbf{Code reuse attacks:} \label{lab:code reuse attacks} Redirects a program’s control flow to execute existing executable code sequences, or \emph{gadgets}, in order to perform malicious operations. These attacks are highly flexible and can be used to bypass common protections, e.g.\ DEP and W$\oplus$X,\footnote{Write XOR Execute (W$\oplus$X) is a security feature that ensures memory regions are either writable or executable, but not both simultaneously.} by using executable code already present in memory.
 \end{enumerate}

A common example of a code-reuse attack is return-oriented programming (ROP)~\cite{RN152}, which relies on the manipulation of return instructions to form a sequence of gadgets.  Each gadget typically concludes with a return instruction; the next gadget in the sequence starts based on the tampered return address of the previous one. It is known that ROP attacks can achieve Turing-completeness~\cite{homescu2012microgadgets}.  Jump- (JOP~\cite{bletsch2011jump}) and call-oriented programming (COP~\cite{carlini2014rop,sadeghi2018pure}) attacks are two other code reuse attacks, which involve tampering forward control-flow transfers, namely indirect jumps and indirect calls respectively, in order to chain gadgets.

\subsubsection{Non-Control Data Attacks} \label{Non-control-data attacks}

Rather than directly modifying control data, e.g.\ function pointers, \emph{non-control data attacks} rely on the manipulation of more general data to indirectly change a program's control flow~\cite{chen2005non,RN155,ispoglou2018block}. Detecting non-control-data attacks is challenging because they traverse legitimate edges of the program's CFG, even though malicious functionality ensues as a result. Let us consider a motivating example. An adversary may execute arbitrary commands on a Linux system if they can influence the arguments to a \texttt{execve()} call in the target program. The target program's CFG is respected, even though the resulting behaviour is unauthorised. Non-control data attacks can be realised in the following ways: 

\begin{enumerate}[start=3,leftmargin=1cm,label={(A.\arabic*)}]
     \item \textbf{Direct data manipulation (DDM) attacks:}\label{attack:decision-making} Refers to attacks that maliciously modify the data used in decisions that affect a program's control flow.
\end{enumerate} 

Branch variable attacks are an example of \ref{attack:decision-making}, where an adversary influences data used in determining the outcome of conditional branches, i.e.\ in `if' and `switch' statements. The resulting data causes another \emph{legitimate} branch to be taken, which nevertheless results in unintended behaviour.\footnote{It is important to recognise this semantic gap: just because legitimate edges of a CFG are followed does not mean that unintended or unauthorised behaviour more abstractly cannot be performed.} An adversary may also attempt to modify the data used in terminating loops, e.g.\ `while' and `for' statements, which we refer to as a `loop variable attack'.

A generalisation of this concept has recently emerged. Data-oriented programming (DOP), first presented by Hu et al.~\cite{RN155}, relies on chaining existing data gadgets in the data plane in order to perform malicious operations. The method is capable of achieving Turing-completeness~\cite{RN155}. Since these flows are part of the program's intended behavior, they do not trigger CFI violations that would be detected using conventional protections. Ispoglou et al.~\cite{ispoglou2018block} later introduced block-oriented programming (BOP) to assist with the automation of data-oriented attacks by identifying and using whole BBLs as data plane gadgets.

\begin{enumerate}[start=4,leftmargin=1cm,label={(A.\arabic*)}]
     \item \textbf{Data-oriented attacks:}\label{attack:dop} Executes a sequence of existing data gadgets on attacker-controlled inputs to perform malicious actions while respecting a program's CFG. 
\end{enumerate}  

\begin{figure}
    \centering
    \includegraphics[width=\linewidth]{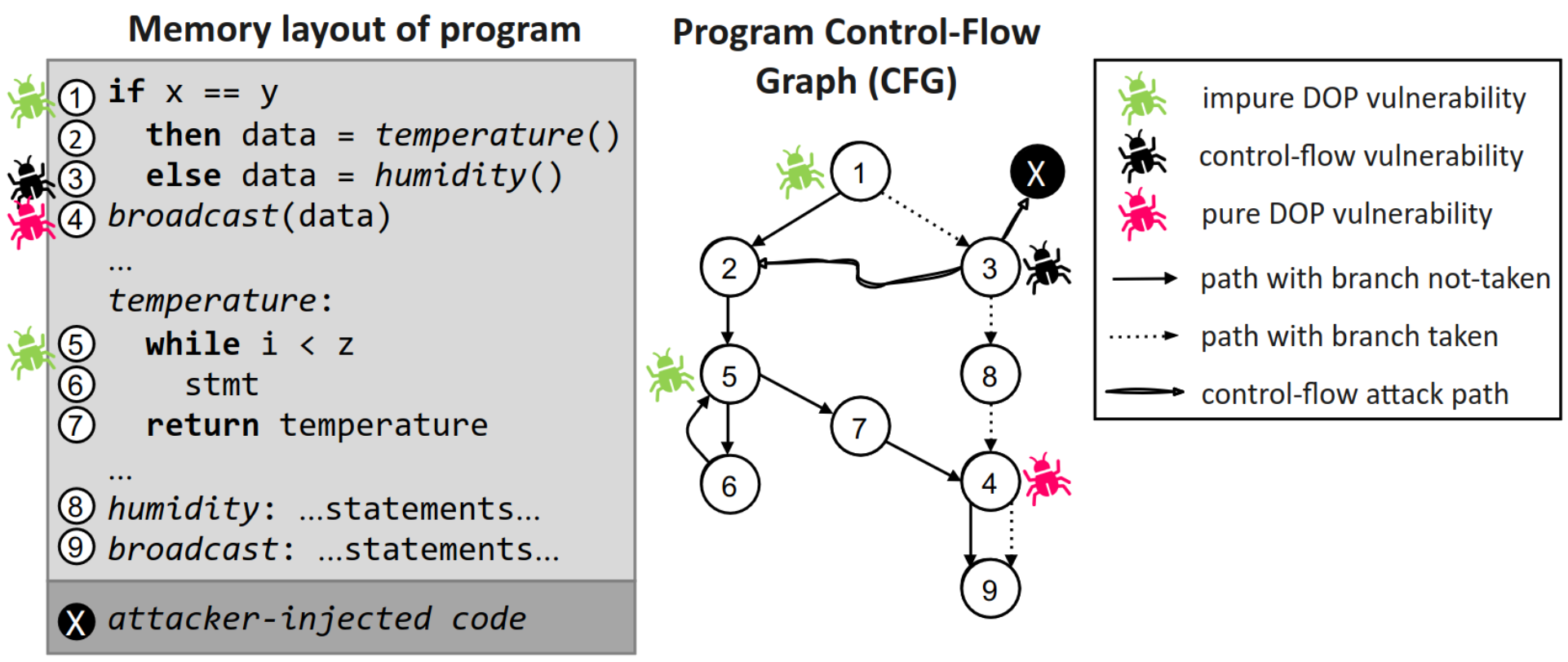}
    \caption{Some control- and data-flow target points from Debes et al.~\cite{debes2023zekra}. An `impure DOP vulnerability' corresponds to \ref{attack:decision-making} if \texttt{x} or \texttt{y} is manipulated to enter the program's `if' block in Line 1. Line 3 may be susceptible to code reuse attacks via code injection \ref{lab:code injection attacks} to divert execution ($3 \to 8$) to ($3 \to X$) or ROP \ref{lab:code reuse attacks} to achieve ($3 \to 2$).  Line 4 has a `pure DOP vulnerability' \ref{attack:dop} if the argument to \texttt{broadcast()} can be influenced.} 
    \label{fig:zekra-cf-attacks}
\end{figure}

In Figure~\ref{fig:zekra-cf-attacks}, we give some common points in which control- and data-flow attacks may be used. In the literature, adversaries with the ability to perform data-oriented attacks are considered to be highly sophisticated~\cite{RN155,LiteHAX}. We emphasise that, if no additional context-sensitivity is introduced, or additional checks are performed on critical data variables, then it is possible for non-control-data and DOP attacks to bypass control-flow integrity mechanisms.

\subsection{Control-Flow Transfer Types}
\label{sec:control-flow-types}

In CFI and CFA, monitoring different \emph{control-flow transfer types} is used to detect particular unauthorised paths taken by a program. It is often undesirable to monitor \emph{every} possible control-flow transfer due to the significant performance overhead that this can occur, especially on constrained devices.  Indeed, not all control-flow transfers are relevant to the security of a system, and it is intuitive to monitor transfers that are common targets for particular exploits of interest. For example, one may wish to pay particular attention to indirect jumps, indirect calls, and return in order to address ROP and JOP attacks. In Figure~\ref{lst:x86-control-flow-types}, we give some examples of such transfers used by X86-64 platforms.\footnote{X86-64 is chosen for illustration purposes; similar examples apply on other instruction set architectures, including ARM and RISC-V. We observe that measurements in CFA proposals are modelled as BBLs or control-flow transfers, rather than specific assembly instructions.} We now describe the control-flow transfer types which have been used in existing work:

\begin{figure}
\includegraphics[width=\linewidth]{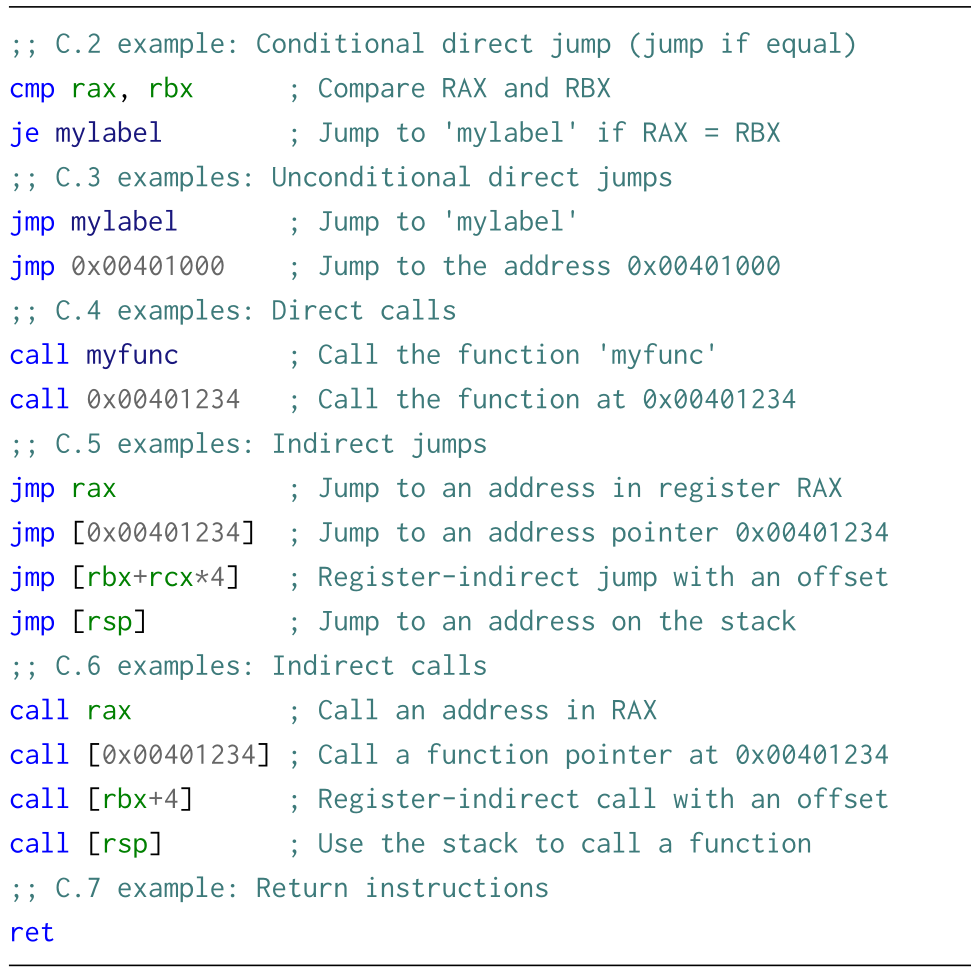}
\caption{X86-64 assembly examples of control-flow transfers.}
\label{lst:x86-control-flow-types}
\end{figure}

\begin{enumerate}[start=1,leftmargin=1cm,label={(C.\arabic*)}]
\item \textbf{All control-flow transfers:} \label{lab:all-control-flow-transfers} This is the most expansive definition. We assume that $\mathcal{C}$ is a maximal set, or has the greatest cardinality, for the target architecture. In practice, all possible control-flow instructions are examined---conditional jumps, direct calls, indirect calls, etc.---imposing the greatest overhead during monitoring and measurement.  It is also the most invasive, requiring the greatest level of instrumentation of the target program (see~\S\ref{sec:instrumentation}). Monitoring all control-flow transfers has been used in many current proposals~\cite{RN144,RN157,LiteHAX,RN149,debes2023zekra,ben2022nanovised,BDFCFA,ARCADIS}.

\item \textbf{Conditional direct branches/jumps:}  Includes all branching and jump instructions that alter the sequence of executed instructions based on a condition. If the condition is true, then the PC is set to a specified address from which it continues execution.  In direct jumps, the target address or label is explicitly stated in the instruction. The address is fixed at compile time.
\label{lab:conditional-direct-jump}

\item \textbf{Unconditional direct branches/jumps:} Transfers the control flow to a specified address or label without any condition using a jump instruction. \texttt{jmp} is the fundamental mechanism for performing unconditional direct jumps in X86-64 assembly.\label{lab:unconditional-direct-jump} 

\item \textbf{Direct calls:} \label{lab:direct-calls} A direct call is used to invoke a subroutine or function by specifying the name or address of the subroutine in the instruction, e.g.\ \texttt{call} (X86-64).
\item \textbf{Indirect branches/jumps:} \label{lab:indirect-jumps} Indirect jumps allow the program to transfer control to an address that is computed at runtime, rather than a fixed address specified in the code. This can include jumping to an address stored within a register, e.g.\ \texttt{jmp rax}; on the stack; or at an address pointed to in memory.
\item \textbf{Indirect calls:} \label{lab:indirect-calls} Similar to indirect jumps, indirect calls invoke a function whose address is computed at run-time or stored in a register or memory location.
\item \textbf{Return instructions:} \label{lab:return-instructions} Used at the end of a function to return to the location of the PC before the function was invoked. On X86, the instruction \texttt{ret} pops the top value from the stack, which should be the return address pushed onto the stack by a previous call instruction, before jumping to that address. 
\end{enumerate}

Some proposals apply different CFI policies depending on the location of the code being examined with the aim of reducing performance overhead. For example, by dividing code into critical and non-critical functions through manual policy specification or using statistical and machine learning-based models~\cite{RN145,RN158,gonzalez2024lightfat,chilese2024one}. For example, ARI~\cite{RN147} divides code into critical and non-critical compartments based on a policy set. All control-flow transfers are monitored except unconditional direct jumps within critical compartments, but transfers within non-critical areas are not considered. OAT~\cite{oat} attests partial code regions that are specified (annotated) by programmers. Within the attested code, conditional direct jump, indirect jump, indirect call, and return instructions are covered. ACFA \cite{caulfield2023acfa} monitors all control-flow transfers within pre-determined code regions including any transfer from outside to inside the region. Other approaches monitor all control-flow transfers within areas that process sensitive \emph{data}, rather than denoted code regions (see DIAT~\cite{diat}). Tracking load/store operations on critical data in memory has also been used, thus bridging the areas of control flow and data flow, but this remains a nascent area of research~\cite{LiteHAX}.%

\begin{enumerate} [resume,leftmargin=1cm,label={(C.\arabic*)}]
\item \label{lab:mixed-cf} \textbf{Mixed:} Different granularity CFI policies are applied depending on code locations or developers' annotation of code to be attested, i.e.\ different control-flow transfers may be monitored accordingly.
\end{enumerate}

\section{Introducing Control-Flow Attestation} \label{sec:introducing-cfa}

A control-flow attestation scheme can be considered as $\mathscr{C} = (T_\mathcal{P}, \mathcal{C}, \mathcal{M}, \mathcal{R}, V)$. The target program, $T_{\mathcal{P}}$, is executed on the proving device, $\mathcal{P}$. The terms `attested program'~\cite{cflat}, `attestation program'~\cite{RN132} and `selected program'~\cite{gedenrasmussen} are used synonymously in the literature.\footnote{Terms like `program,' `target,' or `application' are also used interchangeably, which may be adequate where \prover only executes a single program, e.g.\ microcontroller units.} The proving device may execute other programs during attestation, and so we adopt the term `target program' from \cite{LiteHAX} for clarity. 

$\mathcal{C}$ denotes the control-flow transfer types (e.g.\ direct jumps, indirect calls) being monitored, which were presented in \S\ref{sec:control-flow-types}. A general requirement is that a CFA scheme must be able to monitor unique information about a BBL associated with a single control-flow transfer type. During collection, this information may include the addresses of a BBL's first and final instructions; the control flow transfer's source and destination addresses, which delineates different BBLs; a unique label identifying the BBL;\footnote{IDs are used to overcome issues with Address Space Layout Randomisation (ASLR) in which instructions' memory addresses change with each program execution. Using only instruction addresses, for example, cannot delineate between different BBLs in this scenario. To solve this challenge, the $T_\mathcal{P}$ is instrumented to insert BBL IDs to mark the beginning of BBLs, which do not exist in the original target program.} and the type of control-flow transfer and its complete instruction representation, e.g.\ as machine code.

$\mathcal{M}$ is the facility used to create measurements of the control-flow followed by $T_\mathcal{P}$. We cover measurement techniques, such as cryptographic hashes of CFG nodes, in \S\ref{Measurement}. $\mathcal{R}$ is the procedure for reporting the set of measurements collected by $\mathcal{M}$ to the verifier, \verifier.  This procedure usually involves encrypting and signing collected control-flow measurements, forming an attestation report. Lastly, $V$ is the verification procedure used by \verifier in order to determine whether the measurements correspond to those which are expected, i.e.\ a known, authorised run taken by $T_\mathcal{P}$. A binary decision model is assumed here: if the verification procedure fails, then it indicates the presence of adversarial behaviour and that one of the attacks described in \S\ref{sec:cfa-attacks} was performed. Typically, $T_\mathcal{P}$, $\mathcal{C}$, and $\mathcal{R}$ are run on \prover, while $V$ is run on \verifier. $\mathcal{M}$ is usually run on \prover, but it may be used by both entities.

\begin{figure*}
    \centering 
    \resizebox{\linewidth}{!}{
    \begin{forest}
  for tree={rounded corners, top color=white, bottom color=gray!10, edge+={darkgray, line width=1pt}, draw=darkgray, align=center, anchor=children},
  before packing={where n children=3{calign child=2, calign=child edge}{}},
  before typesetting nodes={where content={}{coordinate}{}},
  [Control-Flow\\Attestation
    [Prover-Verifier\\Paradigm
      [Conventional\\Attestation]
      [[Continuous\\Reporting]
      [[Local Verification\\with Remote Reporting]]
      [Verifiable\\Computation]]
      [Collective\\Attestation]
      ]
    [Trust Anchor
    [Software [Kernel\\Space] [User\\Space]]
    [Hardware
    [TEEs
     [Arm\\TrustZone]
     [Intel\\SGX]
    ]
    [Discrete\\Modules, l*=1.35]
    [ROM]
    [CPU    
    [Custom\\Extensions]
    [Trace\\Modules]
    [PMU]
    ]]
    ]    
    [Instrumentation\\Method
     [Binary\\Rewriting]
     [[Compiler [LLVM] [GCC]]]
      [Dynamic
      ]
    ]
    [Measurement\\Technique
      [Hash-\\based]
      [[Abstract\\Execution]
      [Hybrid]
      [Path\\Log]]
      [Whitelist-\\based]
    ]
    [Adversarial\\Behaviour\
        [Control Data \\Attacks
            [Code\\Injection]
            [Code\\Reuse
                [COP] [JOP] [ROP]
            ]
        ]
        [Non-control\\Data Attacks
            [DDM
                [Branch\\Variables]
                [Loop\\Variables]
            ]
            [Data-oriented\\Attacks
                [DOP]
            ]
        ]
    ]
  ]
\end{forest}
}
    \caption{Knowledge areas of state-of-the-art control-flow attestation schemes.}
    \label{fig:taxonomy_scope}
\end{figure*}
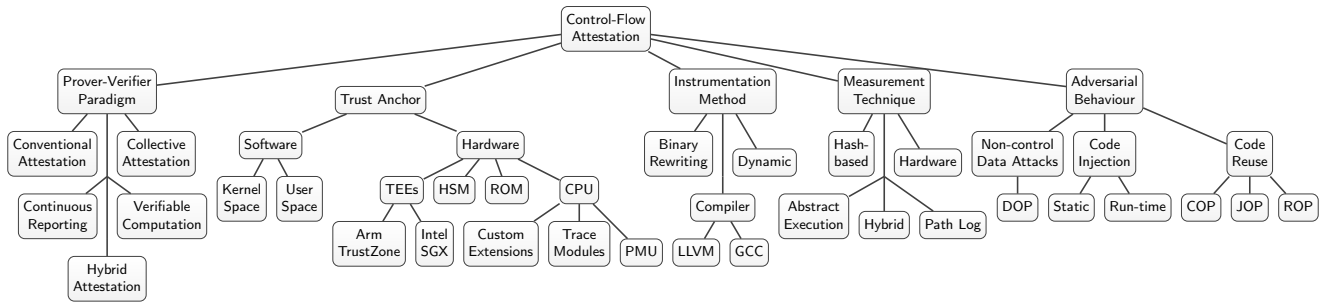

In  Figure~\ref{fig:taxonomy_scope}, we illustrate the knowledge areas that characterise the core elements of CFA schemes. We posit that schemes can be delineated according to their attestation paradigm; the trust anchor; the instrumentation method for collecting measurements; the measurement method; and, finally, the adversarial model under consideration. The coming sections describe the proposals and applications within each area.

\section{Prover-Verifier Paradigms} \label{sec:prover-verifier-paradigms}

We recognise that the majority of CFA schemes operate under an interactive paradigm, which is initiated and verified by a (remote) verifier using a challenge-response protocol. This follows a conventional model of remote attestation used by trusted computing systems, such as the Trusted Platform Module (TPM)~\cite{tpm}. However, other approaches have also been developed in recent years.

\subsection{Conventional Attestation}

 Conventional remote attestation is a two-party interactive protocol between \prover and \verifier, which typically involves an offline and an online phase. The \emph{offline phase} is also known as the \emph{pre-processing}~\cite{RN135} or \emph{bootstrap} phase~\cite{ARCADIS}. Here, \verifier generates the CFG of the target program and identifies and stores all legitimate control-flow paths in a database. \verifier may also modify, or \emph{instrument}, the source code of the target program on \prover to enable the measurement procedure to capture runtime control-flow data. In this model, the proving device belongs to the verifier, who makes these modifications prior to \prover's deployment in the field. On \prover's side, the target program and the measurement and reporting procedures must be provisioned securely, including any key distribution (e.g.\ attestation report signing keys).

The \emph{online} phase involves the measuring and reporting control-flow events. In a typical challenge-response protocol, \verifier sends an attestation request to \prover, which then uses the measurement module to measure control-flow data based on this challenge. \prover may run the target program that accepts the challenge data as its input parameters. When a specific termination condition is satisfied---for example, the final CFG node of $T_{\mathcal{P}}$ has been executed, or a predefined measurement result length has been reached---then the measurement module outputs the result, which is passed to the reporting module. An attestation report is created that is usually integrity-protected and encrypted using one of public-key cryptography, i.e.\ applying a digital signature over the measurement result and metadata, or a symmetric-key method with a shared key between \prover and \verifier. \prover sends the report to \verifier, who decrypts and verifies the report, and performs the check that ensures that the correct control-flow path was followed. Finally, \verifier may perform an action based on this check, such as blacklisting or powering down \prover, or allowing it to access a particular resource.

\begin{figure}
    \centering
    \includegraphics[width=\linewidth]{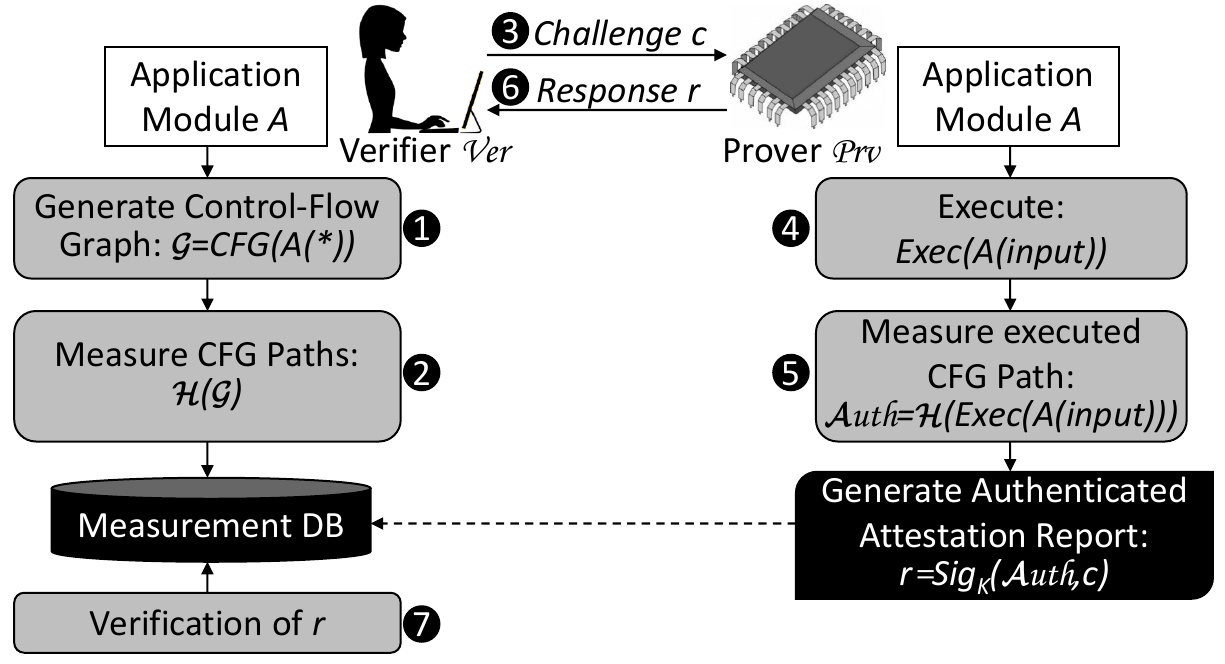}
    \caption{The C-FLAT system model: an example of an interactive CFA scheme~\cite{cflat}.}
    \label{fig:cflat-system-model}
\end{figure}

C-FLAT by Abera et al.~\cite{cflat} is a canonical example of a CFA scheme that follows the conventional attestation model. The proposal begins with an offline phase in which \verifier generates a CFG of the target program in \Circled{1}. \verifier then measures the cryptographic hash values of each path in the CFG in \Circled{2} for determining whether an authorised path was taken. In this scheme, all control-flow transfer types are monitored, i.e.~\ref{lab:all-control-flow-transfers}. Once the device is deployed in the field, the online phase begins. \verifier sends a challenge comprising a program input to \prover in \Circled{3}. \proverprogram makes an execution run based on this input in \Circled{4}. A measurement module, protected by a trusted execution environment, computes the hash-based representation in \Circled{5} before signing and reporting the measurement to \verifier in \Circled{6}. Lastly, \verifier determines whether an authorised CFG path was followed in \Circled{7} by comparing it with the pre-generated measurements from \Circled{2}. The bulk of existing CFA schemes follow a similar challenge-response remote attestation model~\cite{cflat,RN144,RN157,gedenrasmussen,RN169,RN145,scarr,RN132}.

\subsection{Continuous Reporting} \label{Uni-directional Reporting}

\begin{figure}
    \centering
    \includegraphics[width=\linewidth]{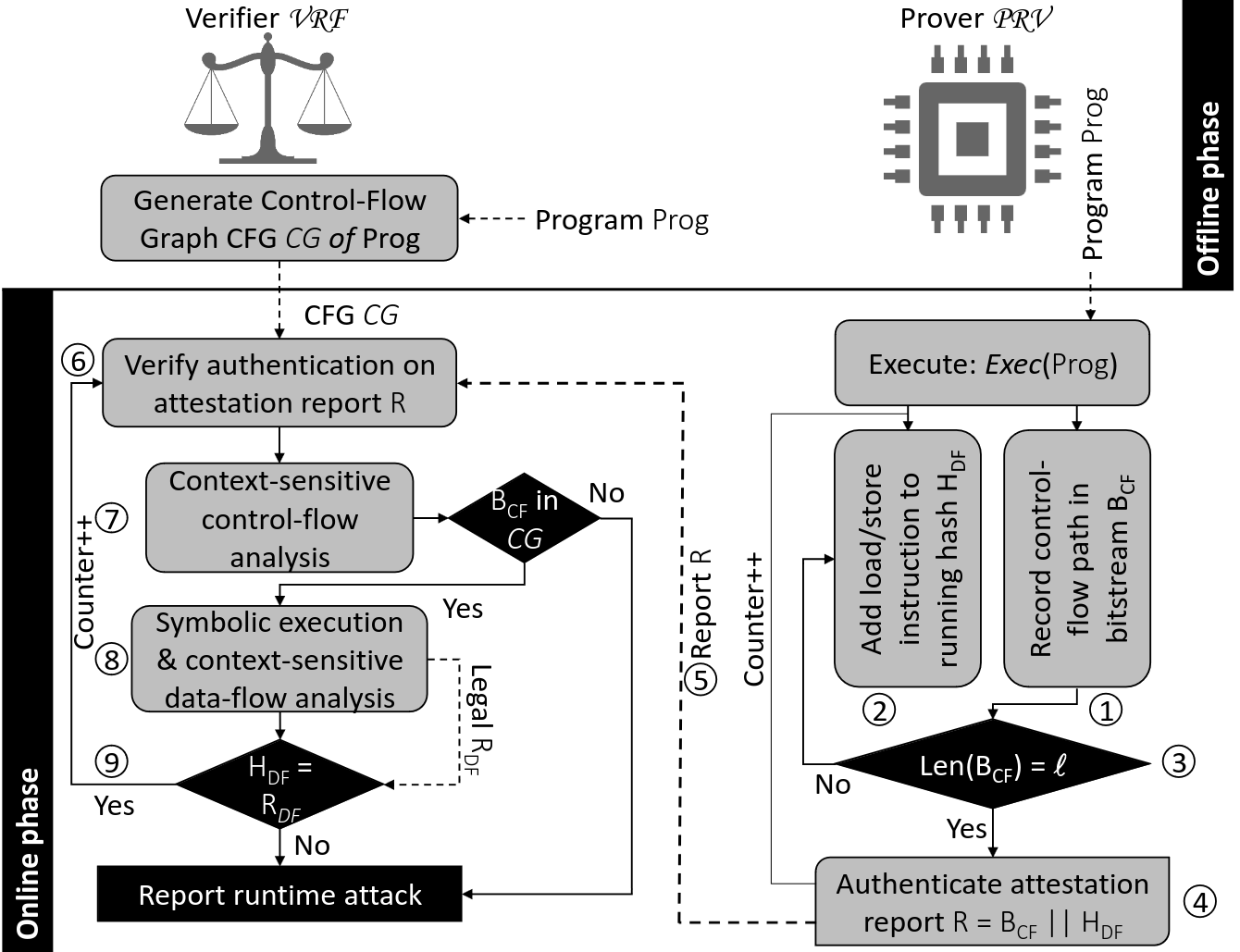}
    \caption{LightHAX: a uni-directional CFA scheme~\cite{LiteHAX}.}
    \label{fig:lighthax-system-model}
\end{figure}

Rather than reporting measurements in a challenge-response protocol, LiteHAX by Dessouky et al.~\cite{LiteHAX} reports measurements continuously \prover$\to$ \verifier. In this work, there is an initial offline phase where \verifier generates the CFG of $T_{\mathcal{P}}$. However, during the online phase, \prover sends attestation reports to \verifier in a one-way, continuous manner. There is no formal challenge-response protocol at run-time. We illustrate this in Figure~\ref{fig:lighthax-system-model}. ACFA \cite{caulfield2023acfa} similarly reports measurements \prover$\to$ \verifier. The transmission of the attestation report is triggered by one of three conditions: (1) the expiration of a timer, which enables $\mathcal{P}$ to send reports periodically; (2) the memory designated for storing measurement results becomes full; or (3) the completion of the attested code execution. Another difference with ACFA is that, after sending the attestation report, \prover pauses the target program's execution until \verifier returns a response to notify whether execution can be resumed.

\subsection{Local Verification with Remote Reporting}
\label{sec:hybrid-attestation}

A separate class of CFA proposals we have identified is those that use a combination of local verification and remote reporting. Geden and Rasmussen~\cite{gedenrasmussen} follows the same offline stage as conventional remote attestation where \verifier extracts a CFG, known as a runtime integrity model, before deploying the proving device. The model is stored internally in a discrete hardware security module (HSM) aboard \prover. The HSM continuously takes control-flow events as input from the CPU in real-time and performs CFI check locally. A remote verifier can acquire a report from the HSM, showing whether any security policy violations were detected over the course of the target program's execution.

CHASE by Dessouky et al.~\cite{RN138} proposes a set of hardware extensions for implementing various run-time security requirements for CFA. Different security modes are supported, where either authorised control flows can be enforced at run-time, or collected for later use in an attestation response. In both cases, it is assumed that valid control flow policies are provisioned on \prover by \verifier prior to its deployment. In remote verification mode, \verifier requests attestation of specific ranges of memory locations at runtime. When local verification is deployed, a measurement is triggered when a local check detects a target attack. When detected, the violating control-flow path is reported to \verifier.

\begin{figure}
    \centering
    \includegraphics[width=\linewidth,interpolate=true]{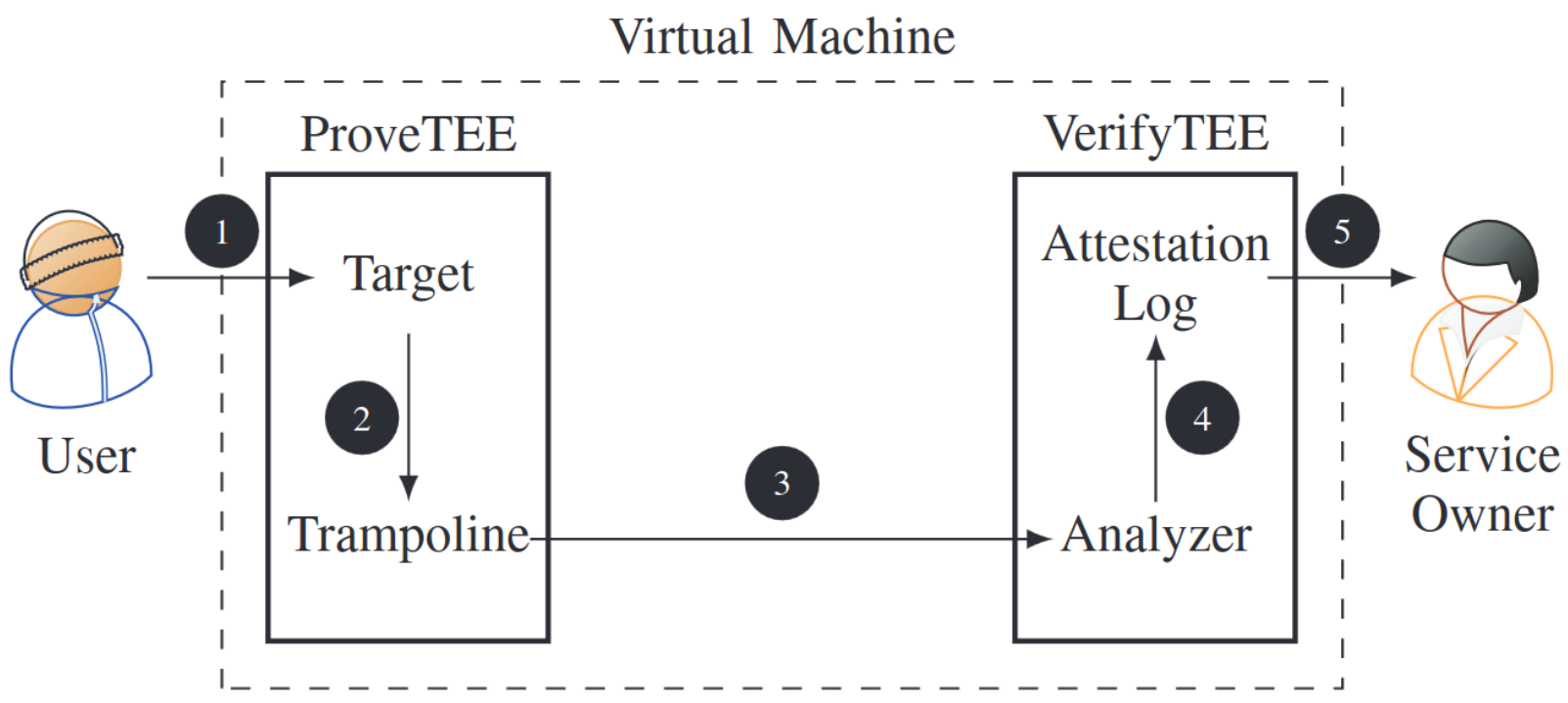}
    \caption{An overview of GuaranTEE~\cite{RN139}. In \Circled{1}, users interact with ProvTEE (\prover), which sends control-flow events to an analyser in VerifyTEE (\verifierprime) in \Circled{2}--\Circled{3}. This data is used in attestation reports for remote verifiers, \verifier in \Circled{4}--\Circled{5}.}
    \label{fig:morbitzer}
\end{figure}

GuaranTEE by Morbitzer et al.~\cite{RN139} investigates control-flow attestation in untrusted cloud environments. The work differs in that \prover and a local verifier, \verifierprime, run on the same cloud-based virtual machine (VM) in Intel SGX enclaves. \prover is assumed to host a security-sensitive application that may be targeted by control-flow attackers; a signing service for \textsc{Covid-19} digital passports is given as an example. \verifierprime measures and analyses valid control-flow transfers, which are passed through a trampoline implemented in \prover. We illustrate this in Figure~\ref{fig:morbitzer}. The target program's instrumentation is performed through extensions to the LLVM compiler framework, allowing \prover's full CFG to be constructed pre-deployment. GuaranTEE supports a form of remote \emph{and} local attestation. At launch-time, each SGX enclave mutually attests the other on the local machine, while enabling a remote \verifier to retrieve attestation reports from \verifierprime.

\subsection{Verifiable Computation}

The traditional remote attestation paradigm for CFA suffers from an inherent problem: it requires that \prover exposes executed paths and other program details to \verifier. This is satisfactory in situations where \verifier can be unreservedly trusted, but it is not tenable where privacy concerns may arise. One example is the attestation of everyday devices, e.g.\ smartphones, owned and used by the general public. Disclosing precise program details could expose privacy-sensitive information about the user's device. To address this problem, Debes et al.~\cite{debes2023zekra} proposed ZEKRA, which developed a new CFA paradigm based on verifiable computation. Here, a third party---an attestation proxy---is introduced in order to perform the attestation verification between $\mathcal{P}$ and $\mathcal{V}$, which are considered mutually untrusted. ZEKRA converts the task into an outsourceable arithmetic circuit through an application of zero-knowledge Succinct Non-interactive Arguments of Knowledge (zk-SNARK). Using this, the proposal enables the proxy to convey only the executed path's correctness to $\mathcal{V}$ without exposing privacy-sensitive program details.

\subsection{Collective Attestation}
\label{sec:collective-attestation}

Conventional CFA assumes a single proving device and verifying entity. However, generalising control-flow attestation to a system or network of devices topology can also be useful. In other words, \prover is not necessarily a single device, but rather a composition of individually attested devices or services whose trust statuses are assessed as a whole. This is known as \emph{collective} attestation~\cite{RN121}. RADIS~\cite{RN156} investigates this problem in distributed IoT device networks. In such environments, a control-flow attack on one device may adversely affect the services performed by another device. For example, one device may produce data, e.g.\ sensor measurements, that influences the actions of a consuming device. To address this, RADIS uses a service-level flow graph (SFG) composed of individual CFGs corresponding to each proving device within a given IoT service. On request, each prover collects measurements of the various control-flow transfers resulting from a challenge given by \verifier. The final device in the SFG returns an aggregated response to \verifier.

\begin{figure}
    \centering
    \includegraphics[width=\linewidth]{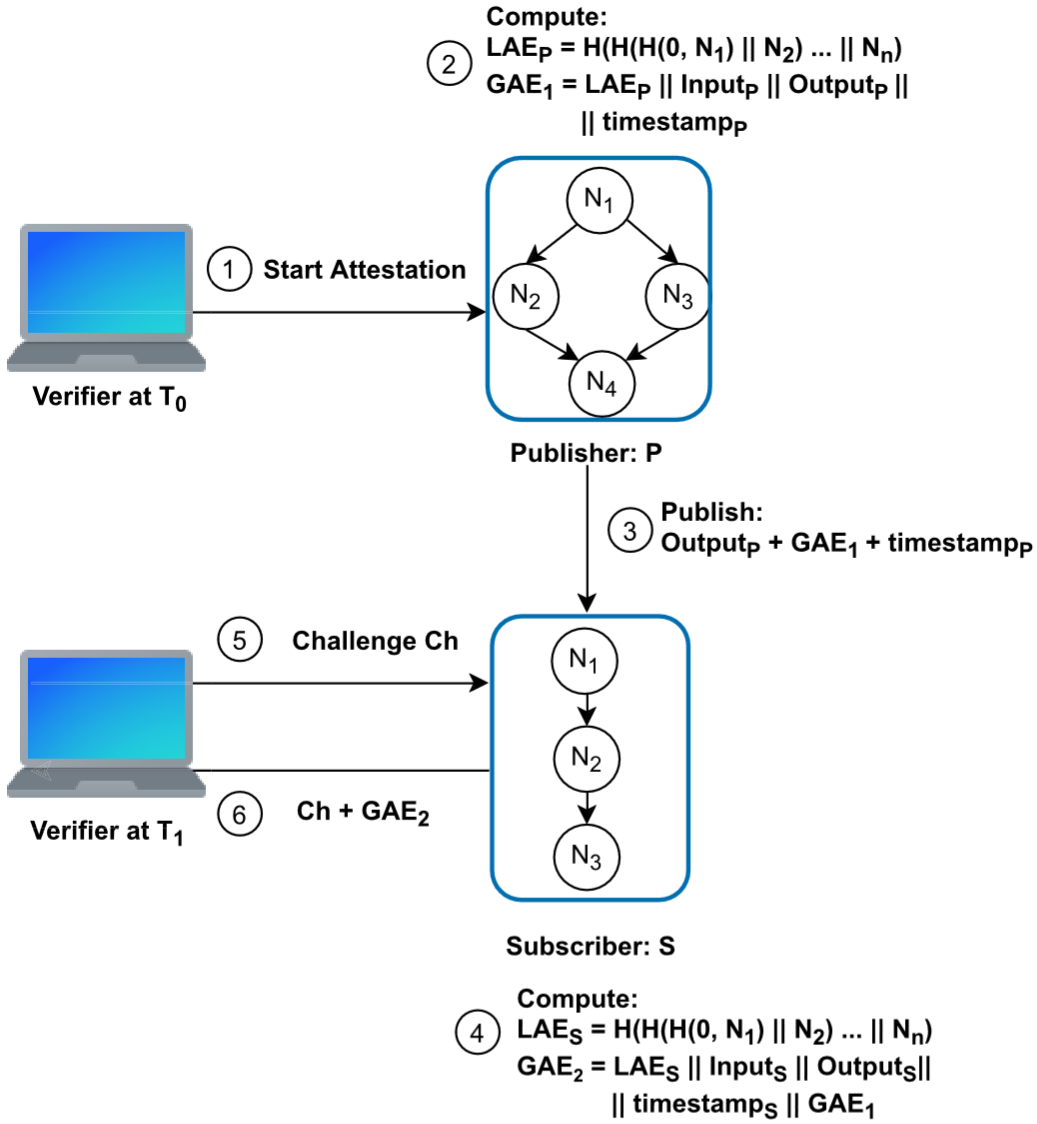}
    \caption{The ARCADIS system model~\cite{ARCADIS}.}
    \label{fig:arcadis}
\end{figure}

Similarly, ARCADIS by Halld\'{o}rsson et al.~\cite{ARCADIS} addresses the attestation of asynchronous distributed IoT services, where multiple devices and services are engaged in publisher-subscriber relationships. In this work, a single \verifier verifies the trustworthiness of multiple IoT devices using hash-based measurements of each node in the devices' CFGs. The system model is given in Figure~\ref{fig:arcadis}. In \Circled{1}, \verifier submits a challenge input to an initial ARCADIS-enabled device, which computes a measurement ($GAE_1$) in \Circled{2}, and sends it to an associated subscriber device, $S$ in \Circled{3}. $S$ computes its own measurement, $GAE_2$, from its CFG in \Circled{4}. Notably, $GAE_2$ contains $GAE_1$ in its output. \verifier makes a separate attestation to $S$ at a future point in time in \Circled{5} in which $GAE_2$ is returned to \verifier in \Circled{6} who checks whether it corresponds to a known `safe' measurement. The scheme can be generalised to multiple target devices; the authors evaluate ARCADIS using a simulated network of up to 250 services.

Li et al.~\cite{RN134} proposed CFRV for attesting networks of peer-to-peer IoT devices, where devices may act as \prover \emph{and} \verifier. The authors address the problem of adversarial verifiers who submit malicious attestation challenges that purposely induce a particular control-flow path resulting in unintended behaviour. To address this, the proposal combines secret sharing and mutual attestation to create an agreed attestation measurement between multiple devices.

\begin{figure}
    \centering
    \includegraphics[width=\linewidth]{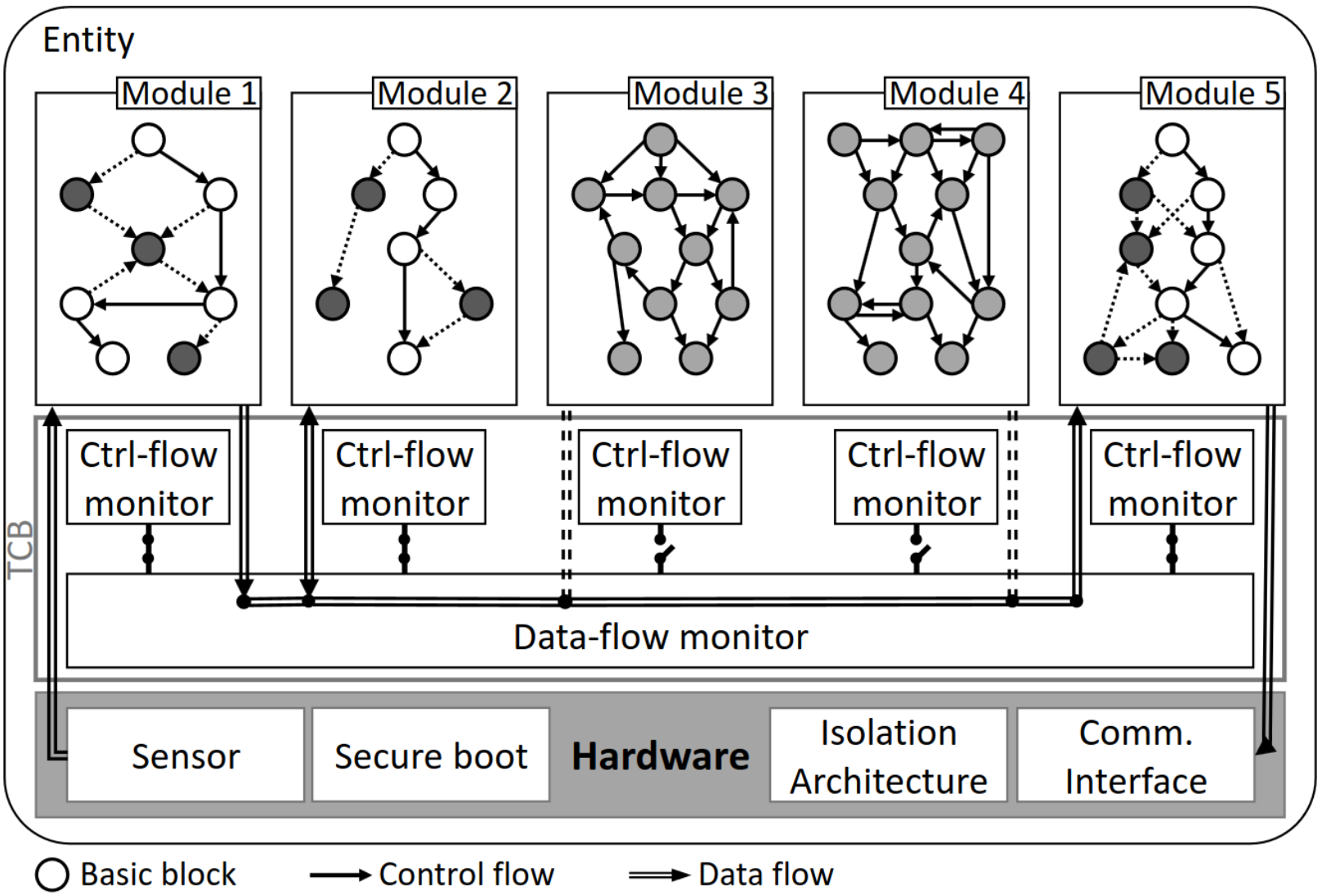}
    \caption{System overview of DIAT~\cite{diat}. Unlike conventional schemes, measurements from each module's CFG are aggregated within attestation reports.}
    \label{fig:diat-overview}
\end{figure}

DIAT~\cite{diat} tackles the attestation of autonomous unmanned aerial vehicles (UAVs) in which several critical components are connected for flight planning, navigation, sensor data acquisition, and actuator control. DIAT provides assurances about the trustworthiness of data flows between such components, which is given in Figure~\ref{fig:diat-overview}, as well as with other UAVs. Let us assume two UAVs, $D_1$ and $D_2$. At a high level, this is achieved by using the data produced by $D_1$ as an input challenge for $D_2$. Each device monitors and measures its own control-flow events using an aggregated hash value, which can be reported to \verifier. 

\section{Trust Anchors and Assumptions for Control-Flow Attestation} \label{sec:trust-anchors}

A \emph{trust anchor} is an authoritative sub-system from which a platform derives its trust; if the anchor is compromised, then so are any dependent components that rely on its functionalities. In theory, trust can be placed in any layer of the platform architecture, from hardened user-mode applications to kernel-mode or hardware-based systems. Nowadays, isolated hardware-based mechanisms are commonly used trust anchors due to their high attack complexity and immunity to powerful (kernel-mode) software adversaries~\cite{abera2016things,shepherd2021physical,koeberl2014trustlite,brasser2015tytan}. Traditional trust anchors of this type include the Trusted Platform Module (TPM)~\cite{tpm}, embedded Secure Elements (SEs)~\cite{reveilhac2009promising}, hardware security modules (HSMs)~\cite{mavrovouniotis2013hardware}, and physically unclonable functions (PUFs)~\cite{schulz2011short}. Modern trusted execution environments (TEEs)~\cite{shepherd2024trusted} use system-on-chip extensions to mitigate the costs of incorporating discrete hardware modules. A number of academic systems have been proposed with similar design goals (e.g.\ TyTAN~\cite{brasser2015tytan} and TrustLite~\cite{koeberl2014trustlite}).

The trust anchor is commonly used to protect \prover's measurement procedure and any key material used in generating attestation reports. For some proposals, this requires that the trust anchor passes traditional static attestation and that it has dedicated memory to store material that only the anchor itself can access~\cite{ARCADIS, RN144,scarr,LiteHAX}. For proposals that rely on TEEs, it is assumed that potentially applicable high-cost attacks cannot be performed by an adversary, such as hardware fault injections and side-channel attacks~\cite{shepherd2021physical}. TEE-based proposals sometimes place trust in certain operating system components, such as kernel modules that initiate the communication with the TEE~\cite{RN149,oat}. In proposals where the kernel serves as \prover's trust anchor, it is sometimes assumed that a method for protecting cryptography keys is available, such as a TPM or a proprietary facility (see Intel MPK in~\cite{RN129}).

Very few CFA proposals are purely based in software, and most systems have some degree of hardware dependency. We observe that CFA schemes can be split into two categories based principally on where security-critical CFA components are implemented on \prover:
\begin{itemize}
    \item \emph{Software-based} systems implement all process components at the software level, typically as libraries in user space and modules in kernel space.
    \item \emph{Hardware-based} proposals rely on a dedicated hardware for performing the measurement and reporting operations. This may be a discrete on-chip security module, an off-chip micro-controller, or a set of bespoke CPU extensions. Alternatively, the proposal uses software whose security is underpinned by a hardware root of trust, such as TEE-based proposals and those that rely on read-only memory (ROM), secure boot~\cite{diat}, or a TPM~\cite{scarr}. 
\end{itemize}

The following assumptions are also commonly made. 

\textbf{Attacks against \verifier are out of scope and it possesses unbounded resources}. State-of-the-art CFA schemes do not consider attacks against \verifier. For example, from malicious insiders, due to supply chain attacks, or any other vector with the intention of undermining the verification ($V$) and measurement procedures ($\mathcal{M}$).  Moreover, it is assumed that \verifier has sufficient computational and storage resources to execute $V$ and, if applicable, $\mathcal{M}$ when required without hindrance.

\textbf{\verifier acts honestly}. It is similarly widespread that \verifier makes no malicious modifications to $T_{\mathcal{P}}$. \verifier deploys the intended program on \prover along with any instrumentation (if applicable) in an honest fashion. It is not considered in the interest of \verifier to instrument code that triggers malicious procedures, issue malicious challenges, or otherwise disrupt \prover after its deployment. (A notable exception is the proposal by Li et al.~\cite{RN134}, discussed previously in \S\ref{sec:collective-attestation}, which investigates collective CFA for peer-to-peer devices).

\subsection{Trusted Execution Environments}
\label{sec:tees}

A \emph{trusted execution environment} (TEE)~\cite{shepherd2024trusted} serves as a common trust anchor in existing CFA proposals. A TEE is a hardened environment in which code and data can be hosted and executed with hardware-enforced isolation. It aims to protect the integrity and confidentiality of code and data at run-time and at rest from kernel-level adversaries in the normal, rich execution environment (REE), such as Android and Linux. Communication between the REE and TEE is achieved through an API that performs the switches between the REE and TEE worlds, and authenticates and authorises the REE application that invokes functions within the TEE.  

Different TEE implementations have been used, but we observe that Intel SGX~\cite{RN158,BDFCFA,RN139} and ARM TrustZone~\cite{RN169,RN132,RN134,diat,yadav2023whole} are widely used choices. Intel SGX follows the enclave model of trusted execution, where applications are partitioned into sensitive and non-sensitive regions. Sensitive portions are initially loaded by the (untrusted) operating system into an enclave, where dedicated CPU extensions provide hardware-enforced access control to enclave-hosted regions when the enclave is executing. The reader is referred to work by Costan \& Devadas~\cite{RN143} for a comprehensive introduction to Intel SGX. 

Intel SGX differs to the trusted world model used by ARM TrustZone, which relies on dedicated operating systems for handling sensitive and non-sensitive applications. We note that TrustZone-A~\cite{RN169,RN132,RN134,yadav2023whole} and TrustZone-M~\cite{diat} have been used, which are compatible with the ARM Cortex-A and -M CPU series respectively. In general, hosting the measurement and reporting procedures in a user-mode TEE application is a common approach; that is, the measurement and reporting procedures execute at S-EL0 in the ARM exception model. Additional extensions to the TEE are also required in a minority of proposals. Notably, Liu et al.~\cite{RN169} use an SRAM-based physically unclonable function (PUF) for deriving device-unique keys used in generating attestation reports. 
Irrespective of whether Intel SGX or ARM TrustZone is used, the same underlying hardware is used to execute normal and secure world applications.

\begin{figure}
\centering
 	\includegraphics[width=0.95\linewidth]{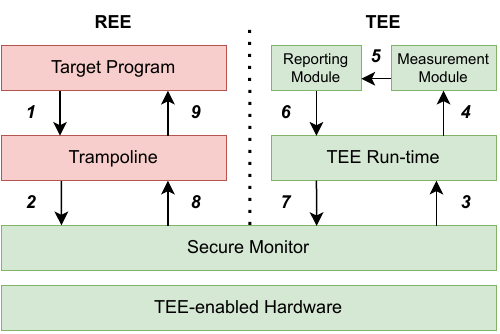}
 	\caption{A simplified TEE-based CFA architecture, showing trusted (green) and untrusted (red) components.}
 	\label{fig:TEE-based CFA architecture} 
\end{figure}

Figure \ref{fig:TEE-based CFA architecture} shows the general architecture of TEE-based CFA schemes using C-FLAT \cite{cflat} as an example. The target program is executed in REE, while the measurement and reporting procedures run within the TEE. In this model, a kernel-based \emph{trampoline} implements the functionality to jump between the normal and TEE worlds. When an instrumented target control-flow transfer is executed, several steps are followed:

\begin{enumerate}
    \item The control-flow event is passed to the trampoline, which gains control of the execution.
    \item The trampoline forwards the control-flow event to the TEE over a secure monitor that performs the world switch to the TEE.
    \item[3--4.] Execution is transferred to the TEE run-time, which forwards the control-flow event to the TEE application (TA) used to host the measurement module.
    \item[5.] The reporting module is invoked immediately after the previous step or after collecting several measurements. The module produces a cryptographic attestation report to be returned to \verifier, typically signed under a hardware- or TEE-bound attestation signing key.  
    \item[6--8.] The reporting module returns the signed attestation report to the TEE run-time environment, which forwards it to the trampoline over the secure monitor.
    \item[9.] Finally, the report is returned to the target program, who then forwards it \verifier over, e.g., a network connection. (The reporting module may directly send the report to \verifier after step 5, depending on whether the TEE can establish network connections itself).
\end{enumerate}

In other work, BLAST \cite{yadav2023whole} partially performs measurement operations in the REE. The proposal uses a register to mark the executed control-flow path in each function, updating the register when target control-flow transfers are executed. The contents of the register are transferred to memory shared between the REE and TEE when the function returns, and then sent to the measurement module in the TEE when the target program finishes execution or the register content reaches a pre-determined length.

We also draw attention to the control-flow property-based attestation (CFPA) scheme sketched by Koutroumpouchos et al.~\cite{RN176}. The proposal seeks to measure higher level behavioural and execution properties of proving programs, i.e.\ as a sequence of abstract states, rather than individual control-flow events. This broadly reflects earlier efforts on property-based attestation (PBA)~\cite{sadeghi2004property}. A system is described in which abstract control-flow events that satisfy particular properties can be used for PBA. The precise control-flow transfers are not specifically defined. A TEE is used as a tamper-resistant trust anchor; however, the described system is not implemented concretely.

\subsection{Normal World Schemes} \label{Non-TEE CFA schemes}
\begin{figure}
    \centering
    \includegraphics[width=\linewidth]{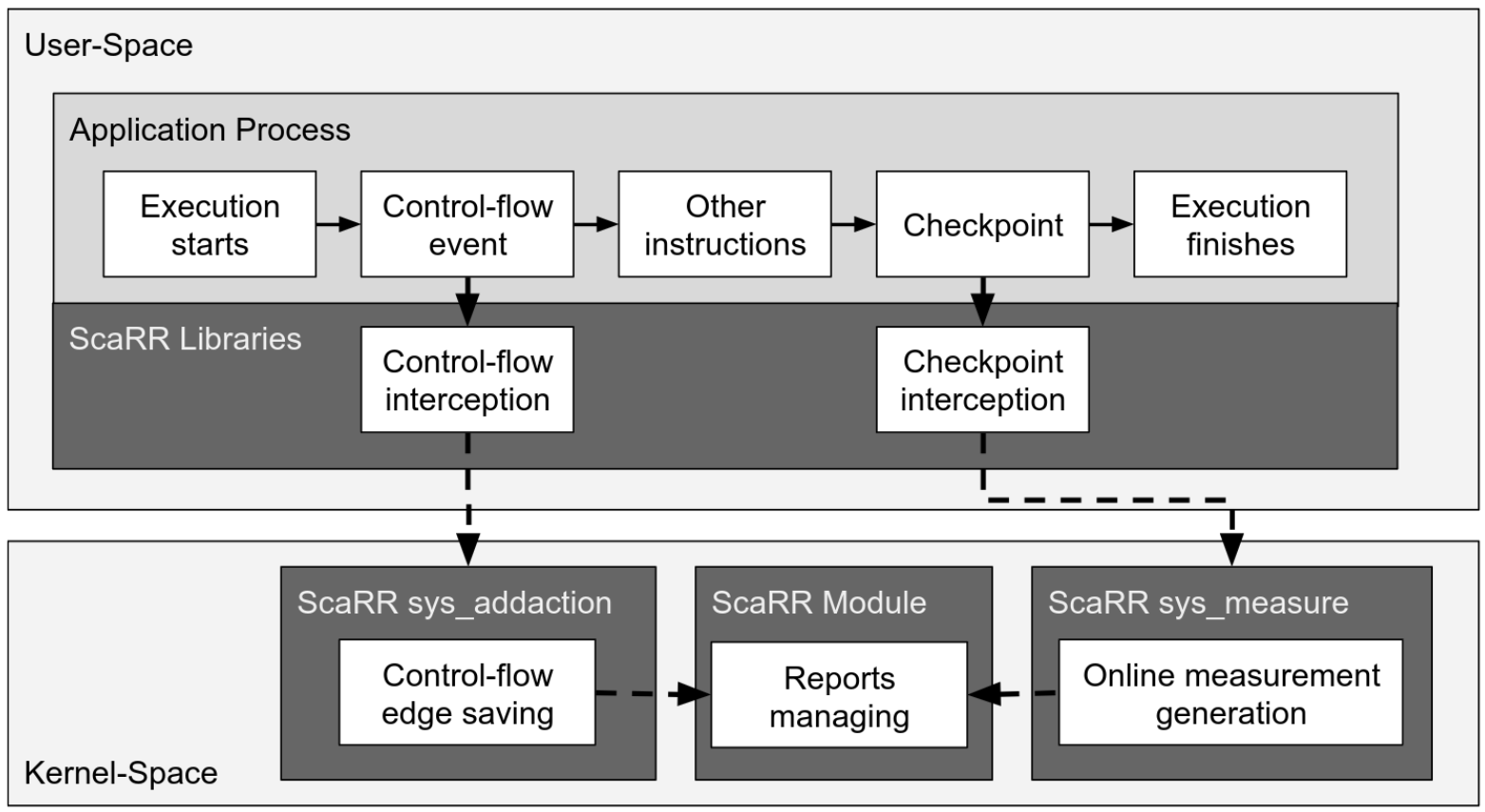}
    \caption{SCaRR's user- and kernel-space components~\cite{scarr}.}
    \label{fig:scarr-overview}
\end{figure}

A small number of proposals deploy the process components, $\mathcal{M}$ and $\mathcal{R}$, in the same execution environment as the target program, whether in kernel- or user-space. This may be considered the `normal' or `non-secure' world, such as Android, Linux or a real-time operating system (RTOS). These schemes do not require a trampoline component to forward control-flow events to the measurement module at runtime. ScaRR~\cite{scarr} relies on kernel-space modules for measuring relevant control-flow events and creating attestation reports. User-space programs hook relevant control-flow transfers, which are passed to the kernel modules through a user-space library (Figure~\ref{fig:scarr-overview}). The design decision was motivated by performance. It is stated that using a TEE (Intel SGX) imposed considerable overhead \emph{``due to SGX clearing the Translation-Lookaside Buffer (TLB)...at each enclave exit. This caused frequent page walks after each enclave call.''} (p.8,~\cite{scarr}). Tiny-CFA \cite{RN160}, a hardware-based CFA scheme designed for low-end microcontroller units (MCUs), makes a similar argument. 

Normal world schemes tend to rely on other security mechanisms to safeguard data against user-space adversaries. ReCFA~\cite{RN129} relies on Intel MPK \cite{intel2016manual}, a proprietary hardware-backed security mechanism for controlling page table permissions from userspace, to protect the measurement module. DIAT \cite{diat} assumes that the prover has access to means of secure storage, secure boot, isolated execution, and inter-process communication. ARCADIS~\cite{ARCADIS} bases its collective CFA scheme on the use of read-only memory (ROM) as a trust anchor for write-protected key storage, encryption and measurement operations. Likewise, Tiny-CFA~\cite{RN160} assumes that attackers are unable to modify any writable memory or access any memory protected by hardware-enforced access control. LAPE~\cite{RN135} relies on standard  memory protection unit (MPU) features to protect the measuring and reporting procedures from unprivileged adversaries.

\subsection{Central Processing Unit Features} \label{CPU-based CFA schemes}

Another common trust anchor is the use of custom or proprietary CPU features. These are predominately used to realise a measurement module that gathers control-flow information while the target program is being executed, i.e.\ at a machine code level. Such CFA schemes can be divided into two sub-categories. The first comprises schemes that employ custom extensions to CPU designs, e.g.\ open-source soft-core CPUs, to capture relevant control-flow events. The second contains proposals that rely on proprietary features of existing CPUs, such as tracing modules that track the history of instructions executed by the processor.

\subsubsection{Custom Extensions}

Dedicated hardware circuits have been used that collect control-flow events directly from processor pipelines and memory controllers. Naturally, this imposes significant invasive modifications to the target platform, but it enables the highest degree of fidelity by monitoring instructions at the closest possible level. A common technique involves reading data from specific points of the processor pipeline, e.g.\ instruction fetch, decode and execute stages. Such proposals are realised as extensions of open-source softcore CPU designs on FPGA implementations, rather than using proprietary CPUs (e.g.\ Intel or AMD). 

\begin{figure}
    \centering
    \includegraphics[width=\linewidth,interpolate=true]{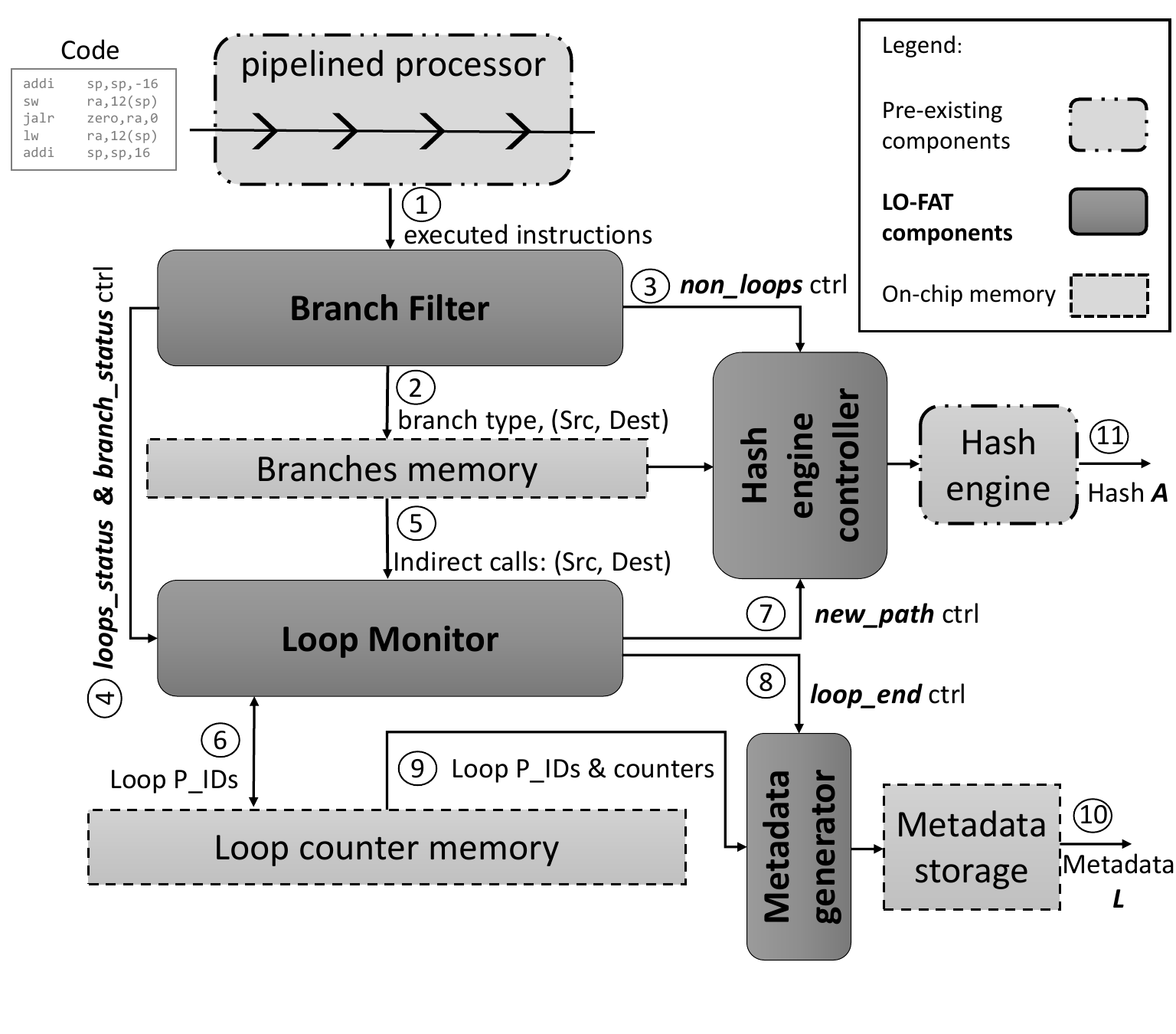}
    \caption{System architecture of LO-FAT~\cite{RN157}.}
    \label{fig:low-fat}
\end{figure}

In this area, LO-FAT~\cite{RN157} hooks the CPU's instruction pipeline in order to measure control-flow events. As shown in Figure~\ref{fig:low-fat}, a branch filter and loop monitor is used to provide inputs to a hash engine and a metadata generator. The hash engine outputs a cumulative hash of the path that was traversed for a given input. The metadata generator outputs data about any loops that were taken, including the loop identifiers, the number of iterations, and the targets of indirect branches that were used.

\begin{figure}
    \centering
    \includegraphics[width=\linewidth]{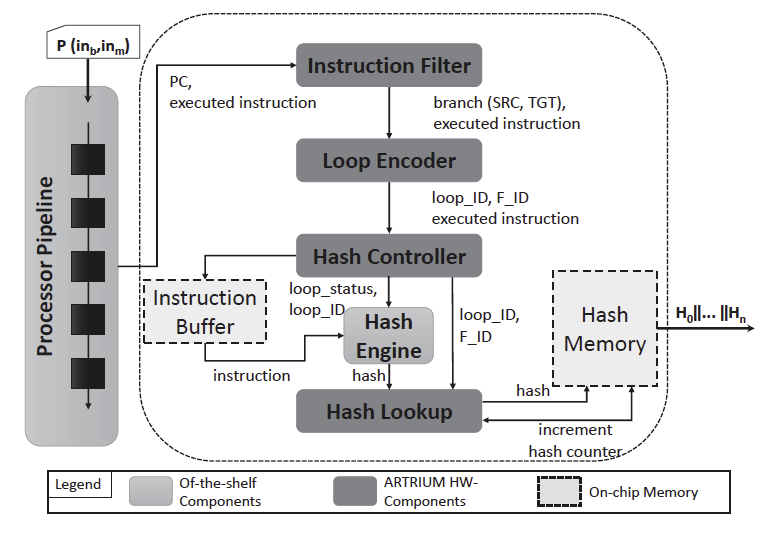}
    \caption{System architecture of Atrium~\cite{RN144}.}
    \label{fig:atrium}
\end{figure}

Atrium~\cite{RN144} (Figure~\ref{fig:atrium}) uses hardware extensions to capture instructions and memory addresses at the execution stage of the CPU's pipeline. The extensions capture the control-flow changes taken by the target program on \prover while executing a challenge from \verifier.  An instruction filter is used to consider only, and all, branch and jump instructions in addition to the current program counter values, and a separate loop encoder is used to track the execution of loops. A hashing engine is used to convert the executed instruction, loop and function IDs---corresponding to nodes of a CFG---into a sequence of cryptographic hashes produced in real-time during the program's execution. These hashes are continuously stored in memory and used in the attestation report returned to \verifier. It is noted that Atrium has reduced on-chip memory requirements while also addressing TOCTOU attacks when compared with LO-FAT.

\begin{figure}
    \centering
    \includegraphics[width=\linewidth,interpolate=true]{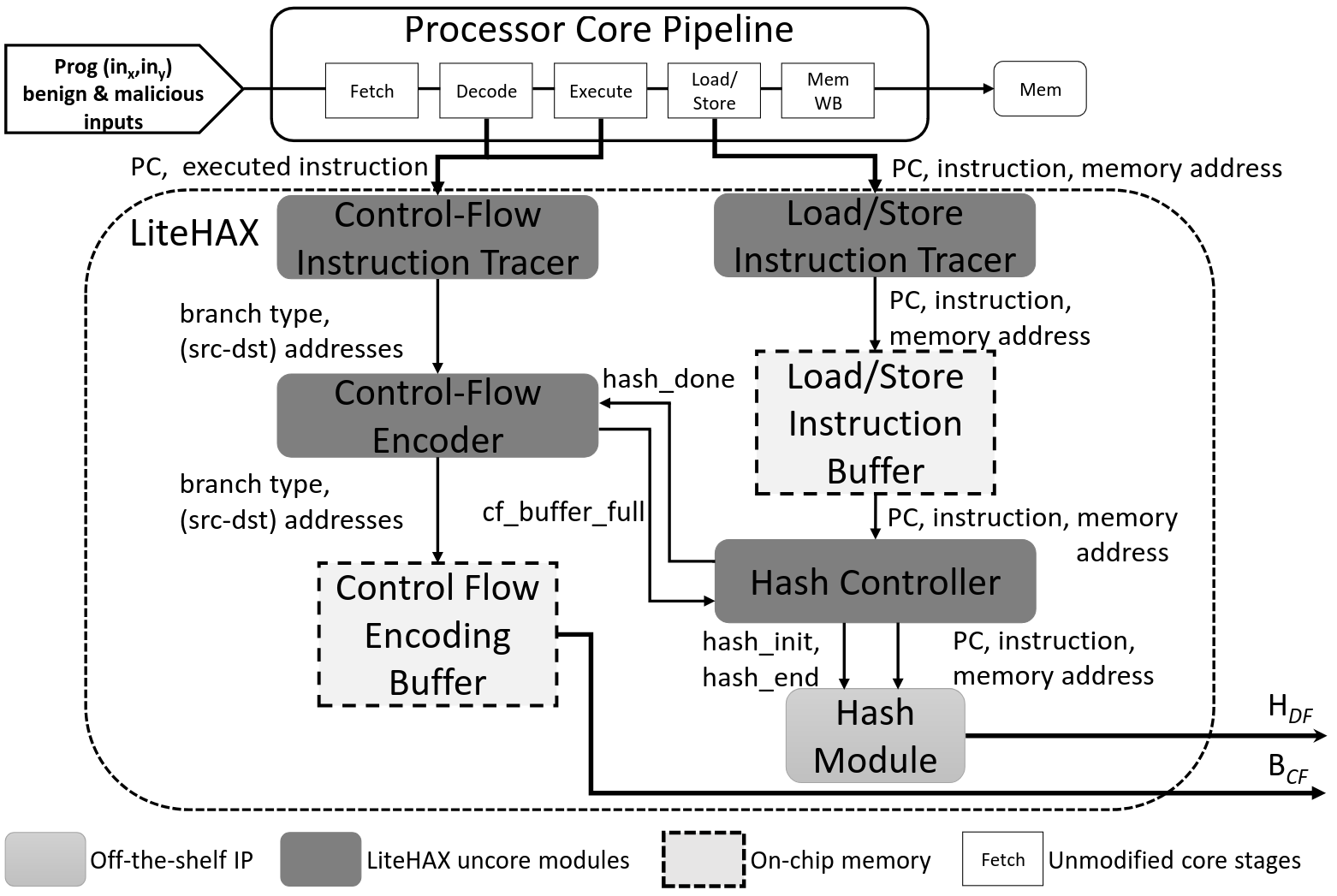}
    \caption{System architecture of LiteHAX~\cite{LiteHAX}.}
    \label{fig:litehax-arch}
\end{figure}

LiteHAX~\cite{LiteHAX} is more complicated in nature, hooking three parts of the CPU's pipeline: the decode, execute, and load/store phases (Figure~\ref{fig:litehax-arch}). The proposal monitors all control-flow instructions \emph{and} memory addresses of load/store instructions in order to address control- and data-flow attacks respectively. A similar hash-based measurement approach is used as Atrium~\cite{RN144} and LO-FAT~\cite{RN157}. However, the LiteHAX hash engine takes encoded control-flow instructions and load/store instructions, their memory addresses, and the current program counter value when calculating the hash values for the attestation report.

CHASE~\cite{RN138} provides a configurable hardware design capable of detecting different adversaries. A selection of up to 10 sub-modules can be used to detect various control-flow attacks, such as code-reuse attacks and DDM attacks. To achieve this, the modules hook into the fetch and decode phases of the processor pipeline, which outputs features to a hashing module that continuously generates hash measurements for attestation reports. Depending on the configuration, both local and remote attestation can be provided. Addressing the strongest adversarial models, e.g.\ control-data and non-control-data attacks, requires the use of 9 sub-modules, whereas the simplest threats requires the use of only four sub-modules.

\subsubsection{Trace Modules}
In recent years, CPU manufacturers have deployed capabilities that actively record control-flow information of programs under execution. This is known as program \emph{tracing} whereby the CPU outputs encoded control-flow information into readable memory. Authorised applications can read this data, also known as a \emph{trace}, to learn the control-flow path followed by a given program. The trace feature is supported by many contemporary CPU processors; for example, Intel CPUs support a tracing module from their 5th generation models, called Intel Processor Trace (PT)~\cite{intel_processor_trace}, which is designed for debugging pruposes. ARM CoreSight~\cite{ning2019understanding} is a comprehensive suite of debug and trace components that are embedded in ARM-based processors, providing developers with method to optimise and debug their applications on ARM architectures. The RISC-V specifications also describe a trace module; however, its availability varies by vendor due to RISC-V's modular instruction set architecture~\cite{riscv_trace_spec,riscv_n_trace}.

Among the proposals we identified, two proposals discuss the application of trace modules for control-flow attestation~\cite{RN164,debes2023zekra}. Papamartzivanos et al.~\cite{RN164} describes a two-tier tracing technique combining a kernel-mode extended Berkeley Packet Filter (eBPF)~\cite{bpf} and Intel PT. By default, eBPF monitors both user space (e.g.\ target program’s configuration and commands executed by the target program) and kernel space (e.g.\ kernel shared libraries and system calls). The data captured by eBPF on \prover is sent to the \verifier for verification. If any violations of specified requirements are detected, Intel PT is used to collect additional control-flow information from the target program. This additional layer of data collection is used to provide assurances about the nature of the detected violation. 

ZEKRA~\cite{debes2023zekra} implicitly relies on trusted components for measurement and reporting with a certified asymmetric key pair in order to provide path authenticity. The proposal does not specify an implementation \emph{per se}, but specifies that the module must chronologically and faithfully provide a trace of the executed path when executing a program region to be attested.

\subsubsection{Performance Monitoring Units}

In 2024, Gonzalez-Gomez et al.~\cite{gonzalez2024lightfat} proposed LightFAt, which relies on a CPU's performance monitoring unit (PMU) as a trusted measurement component. A PMU is a hardware component found on most contemporary CPUs that monitors various hardware events. Such units have been used widely for optimising software performance, identifying bottlenecks, and understanding the interaction with hardware. PMUs can count various types of events, such as the number of instructions executed, cache hits and misses, branch predictions etc., and have been used for numerous security applications~\cite{shepherd2022investigating,xia2012cfimon,maurice2015reverse,zhou2018hardware}. 

LightFAt relies on a CPU's PMU on \prover to monitor the number of instructions per cycles and L1 cache accesses used during the execution of the target program for a given challenge from \verifier. The information about control-flow events is sent to \verifier who then uses a machine learning model to classify the program execution as being malicious or legitimate. The model is assumed to be trained on the target program pre-deployment with known, labelled malicious/legitimate traces. Results of 95\% accuracy and 0.9757 F1-score are presented using a AMD Ryzen 7 2700X 64-bit processor.

\subsection{Discrete Modules}
\label{sec:discrete-modules}

The previous section described several proposals that directly utilise CPU features or hook into the CPU's instruction pipeline using custom hardware extensions. In contrast, several proposals have employed dedicated hardware security modules that are separated from the platform's processor, but are still part of the system-on-chip (SoC). These modules are dedicated to implementing the measurement ($\mathcal{M}$) and reporting ($\mathcal{R}$) procedures in CFA schemes. 

\begin{figure}
    \centering
    \includegraphics[width=\linewidth]{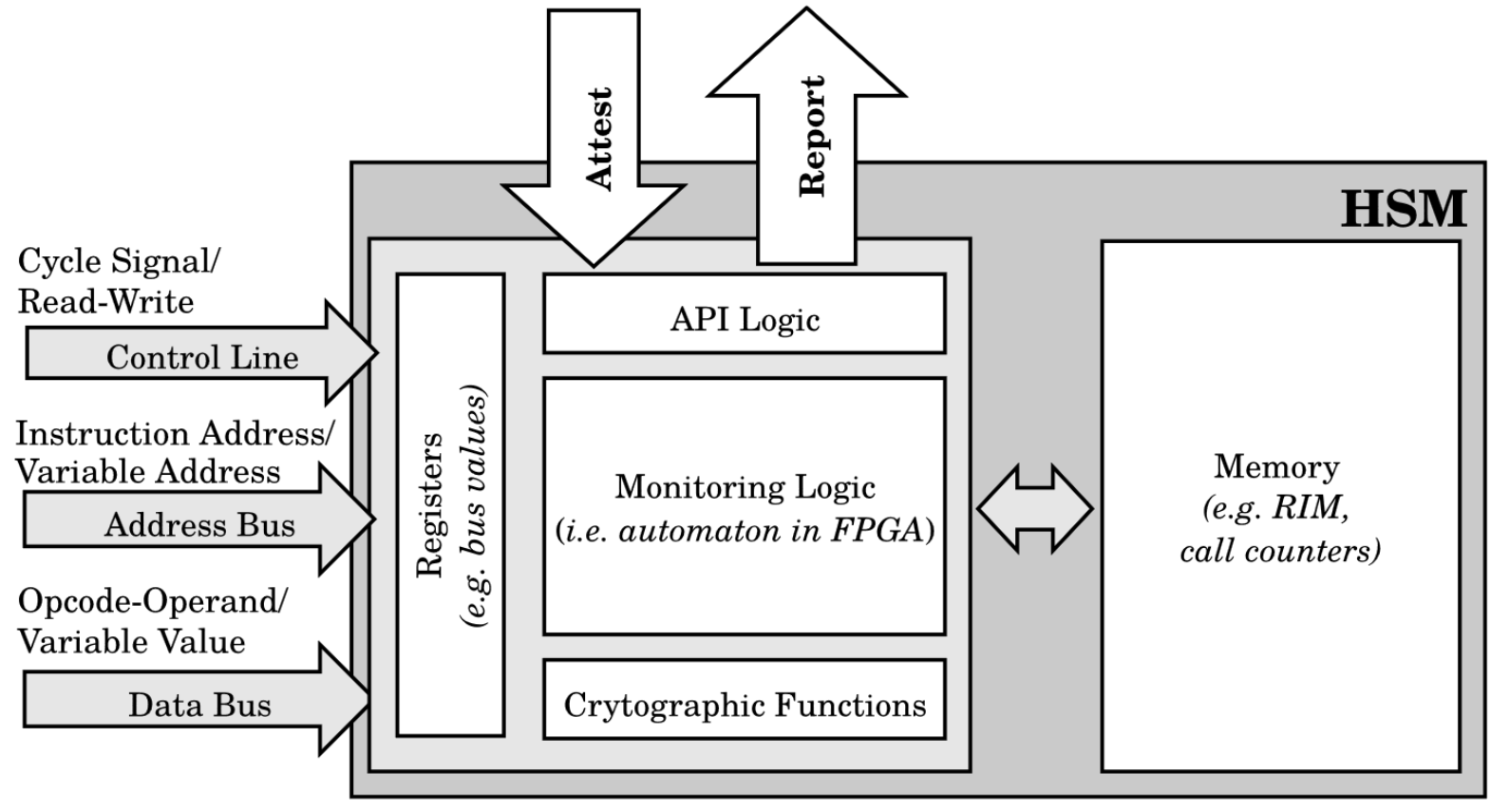}
    \caption{The HSM used by Geden \& Rasmussen~\cite{gedenrasmussen}.}
    \label{fig:geden-rasmussen}
\end{figure}

Geden \& Rasmussen~\cite{gedenrasmussen} proposed a hardware security module (HSM) that collects control-flow information from the system bus (Figure~\ref{fig:geden-rasmussen}).  The HSM continuously takes control-flow events as input from the CPU, including the requested instruction address, opcode, and operand values over the system-on-chip's control line and address and data buses. The collected events are compared with security policies defined in the HSM's runtime integrity model for detecting code injection and re-use attacks. The proposal allows \verifier to acquire a report from the HSM showing whether any security policy violations were detected; it also supports local verification whereby the module outputs a binary result to indicate if a target control-flow hijacking attack is detected. In scenarios where attacks are detected, the module outputs a binary result and provides auxiliary information, such as the last executed instruction. 
 
ACFA~\cite{caulfield2023acfa} proposes a discrete hardware monitor that detects and saves all control-flow transfers during the execution of a target program. A log of these control-flow transfer records, referred to as $\mathcal{C}\mathcal{F}_{Log}$, is written to a dedicated region of read-only memory of a micro-controller unit in a sequential manner. The read-only nature prevents a compromised program from overwriting the event sequence. The module also triggers an interrupt for when the reporting procedure must begin. This launches a trusted measurement routine that computes a MAC over the target program's binary and $\mathcal{C}\mathcal{F}_{Log}$, and sends the result to \verifier for verification.

Tiny-CFA~\cite{RN160} uses a minimalist approach to control-flow attestation, focusing on reducing the trusted computing base and the number of invasive modifications needed to realise a trust anchor. The work employs a simple external hardware module that monitors run-time control-flow states. Tiny-CFA proposes an approach that combines automated code instrumentation with APEX~\cite{nunes2020apex}, which uses a bespoke hardware module to provide a proof of execution (PoX) from a micro-controller unit core.

\section{Instrumentation Methods}
\label{sec:instrumentation}

Collecting control-flow data can be achieved through the process of \emph{instrumentation}. This involves modifying the target program to add instructions that capture, and measure, control-flow events of interest.  Dispatch instructions are used near target control-flow transfers or at the entry of BBLs to this end. When a target control-flow transfer is executed at run-time, the dispatch instructions call pass the transfer(s) to the measurement module for it to be measured. Execution is resumed when the measurement process has completed.

Instrumentation has been achieved in three ways: \Circled{1} at the compiler level~\cite{scarr,RN145,RN135,oat,RN134,RN139,RN147,yadav2023whole}, \Circled{2} by rewriting compiled binaries~\cite{cflat,diat,RN169,RN132,ARCADIS,RN129,RN160,ben2022nanovised,RN149}, and \Circled{3} using dynamic instrumentation~\cite{RN156,RN158,BDFCFA,chilese2024one}. Automated methods are involved in all cases. In theory, manual instrumentation is a fourth possibility---that is, identifying and labelling control-flow transfers by hand---but this is infeasible for non-trivial programs. We recognise that not all CFA proposals require a dedicated instrumentation step. This applies where hardware extensions are used to detect, capture and measure control-flow events within the CPU's instruction pipeline (e.g.\ Atrium~\cite{RN144} and LO-FAT~\cite{RN157}). In the coming sections, we describe methods for instrumenting target programs used by CFA schemes.

\subsection{Compiler-assisted Instrumentation}

\begin{figure}
    \centering
    \includegraphics[width=1\linewidth]{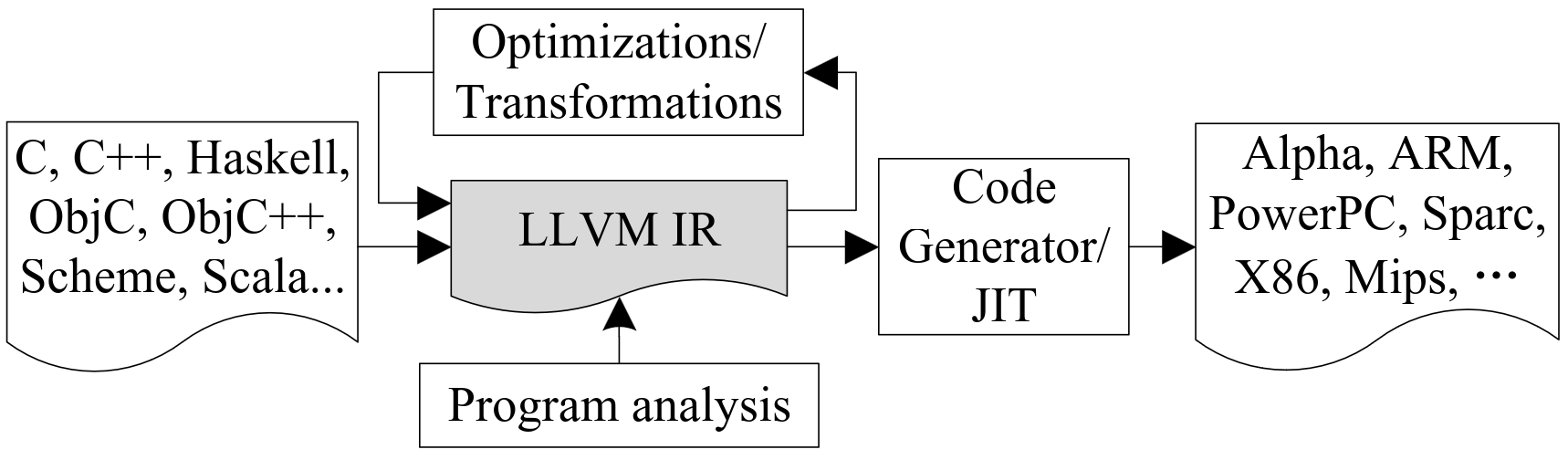}
    \caption{An overview of the LLVM compiler framework~\cite{zhao2012formalizing}.}
    \label{fig:LLVM}
\end{figure}

Compiler-assisted instrumentation assumes access to the program source code, allowing additional code-level semantics to be leveraged during the instrumentation process and enabling precise control-flow tracking and optimisation. Existing proposals tend to rely on the LLVM framework~\cite{RN159} which enables, in theory, the integration of CFA into various input languages and target platform architectures (Figure~\ref{fig:LLVM}). In this vein, the ScaRR system proposed by Toffalini et al.~\cite{scarr} instruments target programs using a bespoke compiler built using the LLVM and CRAB frameworks~\cite{gange2016crab}. The compiler is used to create a `Measurements Generator' for performing several operations. This includes identifying all valid measurements of the program during the offline phase, given as a key-value pairing in Eq.~\ref{eq:scarr}:
\begin{multline}
(cp_{A}, cp_{B},H(LoA)) \Rightarrow [(BBL_{s1}, BBL_{d1}), \\ (BBL_{s2}, BBL_{d2}), \dots (BBL_{sn}, BBL_{dn})]
 \label{eq:scarr}
\end{multline}
Where $cp_{A}$ and $cp_{B}$ are checkpoint identifiers representing an arbitrary sequence of BBLs. $H(LoA)$ is the hashed measurement value of a \emph{list of actions} (LoA), which is the list of CFG edges for traversing $cp_{A} \to cp_{B}$; and $(BBL_{si}, BBL_{di})$ is the source and destination BBL at index $i$.  The compiler is also used for detecting and instrument control-flow events, which are passed to three kernel-space modules responsible for generating measurements. To support this, the Measurement Generator analyses the LLVM intermediate representation (IR) of the target program's processed source code. The CRAB framework is used to generate the CFG itself, which is used to resolve indirect forward jumps.  We note that mapping LLVM to raw assembly BBLs involves the use of dummy code, which is later removed using a custom binary rewriting tool. The authors add approximately 3.5K lines of code (LOC) to CRAB and LLVM 5.0.

GuaranTEE~\cite{RN139} also uses the LLVM compiler framework (see \S\ref{sec:hybrid-attestation}). The authors perform the instrumentation in two LLVM phases: during IR optimisation and the backend stages. In the IR pass, calls are added to the trampoline at the start and end of identified BBLs, i.e.\ at the start and end of BBLs and direct function calls. The backend pass implements trampoline calls before and after indirect function calls, indirect jumps, and before return instructions. Notably, the authors released their LLVM extensions as open-source software. 

BLAST by Yadav \& Ganapathy~\cite{yadav2023whole} is an efficient CFA scheme that relies on a combination of techniques for minimising TEE world switches and the Ball-Larus algorithm for optimal instrumentation placement in target program CFGs~\cite{ball1996efficient}. The authors instrument BLAST as a compiler pass within the LLVM framework in a way that commits CFG path information to a log file on \prover. A TEE is used to sign a hash representation of the log file, which is returned to \verifier on request.

LAPE~\cite{RN135} uses the LLVM-compatible ACES~\cite{clements2018aces} compiler to create and instrument `compartments' on ARMv7-M devices. Programs execute within components where indirect calls and function entry and exit points are recorded and measured using a hash-based approach. We note that, despite LLVM's flexibility, there are some downsides. LLVM is typically not used unchanged: proposals tend to require extensive modifications or extensions of the magnitude of 1,000s LOC~\cite{RN139,scarr,yadav2023whole}.  In practice, this may pose compatibility issues when adapting the schemes to existing toolchains and development environments.

In other work, Hu et al.~\cite{RN145} use a combination of compiler-assisted instrumentation and binary rewriting---discussed in \S\ref{sec:binary-rewriting}---to hook control-flow transfer instructions. The GCC compiler with the \texttt{-finstrument-functions} flag is used to instrument function entry and exit points.  Afterwards, branches within functions are hooked using binary rewriting using the Capstone disassembly framework~\cite{capstone}.

\subsection{Binary Analysis and Rewriting}
\label{sec:binary-rewriting}
Binary rewriting involves making direct modifications to the compiled target program, where control-flow transfer instructions are represented and changed in the form of machine code. This approach can be applied to existing programs, making it suitable for incorporating CFA into software where the original source code is not accessible. The approach has particular applications to retrospectively protecting legacy and proprietary software. 

C-FLAT~\cite{cflat} developed a custom binary rewriting approach for ARM binaries. The authors identify the locations of control-flow instructions, such as indirect branches, which are overwritten with dispatch instructions. The dispatch instructions are used to trigger an assembly routine, which serves as a trampoline for transitioning into ARM TrustZone (see \S\ref{sec:tees}). The Capstone disassembly engine~\cite{capstone} is used to identify control-flow transfers in program binaries to be rewritten with dispatch instructions.

Recording every function call can impose significant performance overhead (see p.\ 3,~\cite{RN129}). To address this, ReCFA~\cite{RN129} derives a CFG from the target program using TypeArmor \cite{TypeArmor} for performing call-site filtering, whereby CFG nodes are skipped where none of its predecessors has more than one successor.  ReCFA uses Dyninst~\cite{dyninst} for the static binary instrumentation of user-space programs on \prover. Dyninst provides a set of libraries and utilities designed for the static (and dynamic) inspection and manipulation of executables. Dyninst enables users to insert, modify, or remove code in a program's CFG without access to its source code. Intel's MPK system is used to protect the integrity of instrumented data structures---the loop stack and path stack---used by control-flow events. The authors compare this to SCaRR~\cite{scarr} where control-flow edges are measured in kernel space, which has a \emph{``potential enormous context switching cost for the control-flow events on an order of magnitude in gigabytes''} (p.\ 6,~\cite{RN129}). The authors combine kernel mode-based functions to preserve code integrity and MPK-protect data structures in order to uphold the integrity of the measurement procedure.

DIAT~\cite{diat}, which attests software modules on drone platforms, realises the instrumentation by making custom extensions to the PX4 open-source flight controller software. Firstly, each software module is assigned a unique identifier for identifying the various flight controller sub-components at run-time. The authors rely on the insertion of dispatch instructions at relevant control-flow events. These instructions redirect execution to a monitor that measures the ID of the software module in question and the source and destination addresses of the control-flow event. The method by which the dispatch instructions are inserted is not clear; however, it is stated that the \emph{``modifications are applied to the code in assembly form''} (p.\ 9,~\cite{diat}).

Liu et al.~\cite{RN169} instrument the target program on \prover at a binary level before it is deployed. A custom binary analysis tool is developed that targets indirect jumps and computes and stores all legal targets in a database held by \verifier. The work instruments the following instructions pertaining to control-flow events (ARM assembly): \texttt{bx lr}, \texttt{pop pc rx}, \texttt{blx rx}, \texttt{bl rx} and \texttt{bx rx}. The instrumentation is used to jump to a trampoline module, discussed in \S\ref{sec:tees}, which sends control-flow data to a measurement module implemented within ARM TrustZone, which includes the values of return addresses and function pointers to be reported to \verifier.

\subsection{Dynamic Instrumentation}

Dynamic instrumentation refers to the method of inserting, monitoring and verifying code into a program while it is running. This compares to instrumenting target programs using static analysis and manipulating the executable before execution (e.g.\ using binary rewriting). Dynamic methods rely on the insertion of probes or tracking mechanisms into a running program. These probes can monitor the control flow of the application in real-time, identifying any deviations from expected paths. 

Common dynamic instrumentation tools used more generally in the area of control-flow integrity use Intel Pin~\cite{luk2005pin}, DynamoRIO~\cite{bruening2004efficient}, and LLVM-based methods~\cite{engelke2020instrew}. In the area of control-flow attestation specifically, GACFA~\cite{RN158} relies on Intel Pin-3.15 to instrument the target program during runtime. Here, control flow events, which are collected using Intel Pin, are sent to the scheme's measurement module running in an Intel SGX enclave. The measurements are calculated using a hash-based approach; the final measurement is signed under a secret key within the enclave and provided within an attestation report. A similar approach is used in the BDFCFA~\cite{BDFCFA} proposal, which uses Intel Pin for binary instrumentation, and presents an approach with bidirectional attestation for blockchain applications. 

RADIS~\cite{RN156} uses the Python \texttt{trace} module to dynamically instrument target functions at the interpreter level. A bespoke implementation of the module in a TEE is recommended to protect the measurement procedure from kernel-mode adversaries; however, the proposal was neither implemented nor evaluated.
\section{Measuring and Monitoring Control Flow} \label{Measurement}
The measurement process is one of the key facets of a CFA scheme. It is at the core of \prover demonstrating to \verifier that an authorised path/walk was followed through the target program's CFG. For a given CFA scheme, the measurement method typically remains constant, regardless of the control-flow paths or the target program in question. However, these methods tend to vary significantly \emph{between} CFA schemes, which we elaborate upon in this section. 

The measurement process measures the \emph{whole} or \emph{part(s)} of the target program's CFG:



\begin{itemize}
    \item \emph{Whole program measurement:} The whole control-flow path used by the target is measured, comprising the nodes traversed by \prover using a challenge supplied by \verifier. The task of \verifier is to determine whether the measurement returned by \prover corresponds to a known, `safe' value reflecting the execution of the target program from start to finish.

    \item \emph{Partial measurement:} Particular aspects of the control-flow path are measured, such as loops, certain branching instructions, functions etc.\ The overall measurement may comprise several such measurements as a proxy that the target program had executed correctly and without the presence of an adversary. Given a measurement result from \prover, then \verifier checks whether these (partial) program measurements conform to known valid results. 
\end{itemize}

A CFA scheme may measure the whole program \emph{and} partial elements, or multiple aspects of partial elements, simultaneously. BLAST~\cite{yadav2023whole} serves as a good example, which uses optimal instrumentation of loops and functions to achieve whole program control-flow attestation efficiently. 

We can also consider measuring techniques as acting at different levels of abstraction. \emph{Path-based} techniques measure an entire or partial control-flow path within the target program's CFG, i.e.\ path-level granularity. In contrast, \emph{control-flow transfer-based} methods monitor individual low-level transitions at each branch, function call, or another relevant control-flow event. There is little information stored explicitly about the path.

At first glance, it may seem that path-based measurements are superior because they consider the entire path explicitly.  However, there can be a significant measurement \emph{range} of potentially legitimate paths for non-trivial programs. Recall that, prior to \prover's deployment, \verifier usually must learn all of the legitimate control-flow paths within an offline phase. This may be stored in a database in the form of <key, value> pairs. (Using hash values to aggregate and represent the nodes traversed by \prover is a common approach~\cite{cflat,RN144,scarr,diat,ARCADIS,RN156,ben2022nanovised,RN134,RN132,RN135}). In the online phase, \verifier issues an attestation challenge, and \prover responds with a measurement that \verifier matches against the records in its database. This method suffers from the \emph{path explosion} problem, where the number of control-flow paths grows dramatically with the number of branches in the target program. The storage requirements placed on \verifier can be enormous.   It is also possible that \prover's CFG may contain non-deterministic behaviours that are not straightforward to calculate \emph{a priori} as a legitimate path, such as branches that depend on input/output (I/O) data.  


In comparison, monitoring and measuring events at a control-flow transfer level can mitigate the challenges above. Rather than tracking all potential control-flow \emph{paths}, only legitimate control-flow \emph{transfers} are stored. For example, using a whitelist or otherwise flagging any anomalous transfers that are encountered. On the down side, this method can impose significant run-time overhead on \prover if many control-flow transfers are monitored and measured during execution. An interesting observation can thus be made: the choice of measurement approach reflects a time-space complexity trade-off between the storage requirements of \verifier and the run-time overhead on \prover.

\subsection{Hash-based Measurements} \label{Hash-based measurement}
A na\"{i}ve method is to instrument the target program on \prover in order to capture every control-flow transfer. When \prover traverses the CFG of a target program, \prover stores a list of these transfers and sends it to \verifier as part of an attestation response. However, such an approach can lead to prohibitively large attestation reports that may be unsuitable for constrained devices. Many schemes, therefore, make use of hash-based measurements to represent control-flow paths in a space-efficient manner.

\subsubsection{Hash Computation}
Let us consider the above example once more. One solution to reduce the memory requirements on \prover is to make a single hash calculation over the collected control-flow transfers and return this (hashed) result to \verifier after the CFG's terminating node has been reached. This method is, again, not  widely used due to the need to store every control-flow transfer before the hash function has been applied. 

To address this problem, hash-based CFA proposals have opted for a cumulative approach to hash computation. The measurement module performs a hash computation at \emph{every} control-flow transfer using data from the previous transfer. This can be described by the recurrence relation in Eq.~\ref{eq:cumul-hash}:
\begin{equation}
H_i = 
\begin{cases} 
h(d, C_i) & i = 0 \\
h(H_{i-1}, C_i) & i > 1
\end{cases}
\label{eq:cumul-hash}
\end{equation}

\begin{figure}
\centering
 	\includegraphics[width=0.45\textwidth]{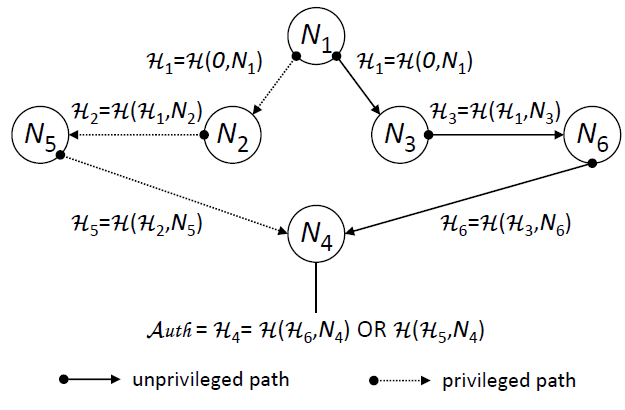}
    \caption{C-FLAT's cumulative hash calculation approach~\cite{cflat}.}
 	\label{fig: an example of cumulative hash calculation} 
\end{figure}

Where $i$ denotes the node number, $H_i$ refers to the hash value generated at node $i$, and $h(\cdot)$ is a suitable hash function. \(C_i\) is the control-flow event at $i$, which may be represented by a (\textit{src}, \textit{dst}) pair \cite{RN157}, an (\textit{entry}, \textit{exit}) pair \cite{scarr}, or a BBL ID \cite{cflat}. $d$ represents the default value used with the first control-flow event when $i=0$, e.g.\ $0$ in C-FLAT \cite{cflat} or \(h(0)\) in GACFA \cite{RN158}. The main advantage of cumulative hash-based approach is that it does not require \prover to store all control-flow transfers. Once a round of calculation is completed, the corresponding control-flow transfers can be discarded. In Figure~\ref{fig: an example of cumulative hash calculation}, we present a canonical example of a cumulative hash-based approach used by C-FLAT~\cite{cflat}, demonstrating the hash values at different nodes of a target program's CFG.

Common hash functions used in existing CFA proposals can be categorised as standard cryptographic hash functions, and multi-set hash functions.

\emph{Standard cryptographic hash functions:} BLAKE2~\cite{RN165} is a common choice for CFA schemes targeted at embedded systems with limited computational resources~\cite{RN138,scarr,oat,RN147}. BLAKE2 is known to significantly outperform ($\sim$50\%) competing functions, such as Keccak and SHA-2, in raw execution time (cycles per byte)~\cite{RN165}. We note that SHA-256~\cite{RN156,RN158}, SHA-512~\cite{ARCADIS} and SHA-3~\cite{RN157,LiteHAX} have also been employed in existing work, albeit to a lesser extent. Worryingly, cryptographically broken algorithms have also been investigated by some schemes (MD5~\cite{RN135,ARCADIS}).

\emph{Incremental Multi-set hash (MSH) functions:} Some CFA schemes have explored specialised hash functions with space-efficient properties. Incremental MSH functions~\cite{RN173} take as a collection, or \emph{multiset}, of arbitrary elements and their occurrences. An MSH function allows one to incrementally generate different permutations of the multiset while enabling the testing of equivalence relations. For example, consider the multiset $\textbf{M} = \{\langle A,3 \rangle, \langle B,2 \rangle, \langle C, 8 \rangle,  \langle D, 1 \rangle\}$, where $\langle A, 3 \rangle$ denotes the element $A$ and $3$ represents its number of occurrences (or multiplicity).  Consider an MSH value constructed from this multiset where $D$ is added to an existing ordered set, $\{A, B, C\}$, i.e.\ $H_1 = \{A, B, C\} \cup D = \{A, B, C, D\}$. Consider another MSH value, $H_2 = \{D, B, A, C\}$. An MSH scheme enables one to verify the equivalence of $H_1$ and $H_2$.

MSH schemes are used by DIAT~\cite{diat} and CHASE~\cite{RN138}. They allow the hash value of the set to be efficiently updated as new elements, i.e.\ control-flow transfers, are added. Incremental MSH schemes are used to efficiently update new control-flow transfers, avoiding the storage or recalculation of the entire set upon adding a new element~\cite{RN173}. Compared to conventional cryptographic hash functions, MSH functions are space-efficient for CFA applications, reducing the number of possible hash values to be stored in a database belonging to \prover or \verifier. 

The space-efficient property of MSH functions can also pose some downsides. Namely, they do not consider the order of control-flow transfers in the set. It is possible that multiple legitimate control-flow paths have the same hash value, thus \verifier loses contextual control-flow path  information. Even if a hash value in the measurement result matches one in the database, \verifier can determine only the number, not the order, of executed control-flow transfers. This could be fatal if an adversary finds a sequence of malicious control-flow transfers that produces the same MSH value as the legitimate sequence. We pose this issue as a research challenge.

\subsubsection{Discussion} \label{Pros and cons}
The main advantage of hash-based measurement is that it can generate short, constant-length measurement results, irrespective of the target program's size or the number of control-flow events. The length of the measurement result is a major factor to influence the attestation report size, and thus the space and network complexity when the report is computed on \prover and sent to \verifier. 

%
However, hash-based measurements present some major disadvantages. The first is the aforementioned path explosion problem: if a target program contains several loops with non-trivial terminating conditions, then it is possible that relatively short programs can impose significant storage requirements on \verifier. Furthermore, if an attack \emph{does} take place, then \verifier cannot determine \emph{how} or \emph{where} any violations occurred in the control-flow path followed by the target program from the hash measurement alone~\cite{RN169}. The presence of adversarial behaviour can be detected, but its precise nature cannot, which may pose difficulties in identifying and patching any security vulnerabilities.  This is due to the inherent pre-image resistance and avalanching effects of cryptographic hash functions: using just one additional control-flow transfer in the hash computation yields a significant difference in the resulting hash value.

A segmented hashing approach has been used to reduce, but not eliminate, the effects of path explosion. Here, the target program is divided into multiple segments, and a separate hash computation is performed for each segment. Atrium~\cite{RN144} considers each loop in the target program as a separate, independent segment. Let us consider a toy target program containing five sequential loops, each of which may iterate up to 10 times. A separate hash computation is performed for each loop. \prover will generate six hash values to represent the entire control-flow path, one for each loop, plus one hash value to represent a control-flow path excluding all loops. Consequently, \verifier must only store 51 legitimate hash values (10 for each of the five loops) and one legitimate value for the control-flow path excluding loops, signfiicantly reducing the storage requirements versus a conventional hash-based approach (resulting in \verifier storing 100,000 control-flow paths). Segmented hash computation has also been used to tackle conditional branches and recursive functions~\cite{scarr}.

\subsection{Other Path Log Representations}

Rather than summarising the control-flow path using hash values, other proposals have used more sophisticated methods for logging and representation. ISC-FLAT by Neto \& De Oliveira~\cite{RN149} aims to address the problems caused by system interrupts during the measuring process. Consider a simplified example in which \prover traverses the nodes $N_A \to N_B \to N_C$ to produce a hash value $H(N_A, N_B, N_C)$ as part of a legitimate control-flow path run on \prover. In realistic scenarios, a software or hardware interrupt may spontaneously cause another sequence of nodes to be visited mid-measurement, i.e.\ $N_A \to N_B \to (N_E \to N_F \to N_G) \to N_C$, where $(\cdot)$ contains the nodes visited by an interrupt handler. Hashing these nodes would produce an unintended hash value, undermining the reliability of the measurement process.\footnote{A na\"{i}ve solution is to suspend interrupts altogether during the measurement process, but this can give rise to wider system issues.} 

ISC-FLAT addresses this problem by using a TEE-based application to monitor the control-flow transfers of an instrumented target program and trapping all interrupt requests within the TEE.  The TEE is used to protect the measuring process from kernel-mode adversaries. When an interrupt is trapped, the TEE application notes the points at which the interrupt service routine was activated, which is reflected in the report sent to \verifier. The system stores path logs of the form $\{A^{i}_{0}, A^{e}_{0}, A^{i}_{1}, A^{e}_{1}, \dots, A^{i}_{n}, A^{e}_{n}\}$, where $A^{i}_{x}$ represents the memory address of the first instruction of node $N_x$ in the program's control flow graph, and  $A^{e}_{x}$ is the destination of a branching instruction in $N_x$. In other words, a sequence of addresses and destinations is used to represent the control flow, rather than aggregating them as a single hash value.  Control-flow abuse is then detected through the inspection of the path log that is sent from \prover to \verifier. 

BLAST \cite{yadav2023whole} uses Ball-Larus numbering to assign a numeric value to each edge of the target program's CFG. Every possible path is marked by a unique integrity identifier---a path ID---which is the sum of the values of the edges included in the path. When a function call has finished execution, the path ID, target address of the function call, and return instruction are recorded and stored in shared memory between the REE and TEE. When a given length has been reached, the log is sent to measurement module within the TEE to compute a cumulative hash value. BLAST can send two different measurement results. By default, \prover sends a single hash value as a measurement result to \verifier. If \verifier cannot match the value in its database, it requests that \prover provides a sequence of tuples \texttt{<func, pathID>}. \prover then computes the Whole Program Path (WPP) representation~\cite{larus1999whole} from the sequence and sends it to \verifier. 

Another example is RAGE \cite{chilese2024one}, which relies on a machine learning-based approach using unsupervised graph neural networks to avoid computing a complete CFG of the target program. In the training phase, \verifier uses one legitimate control-flow path represented as a sequence of BBL addresses as a partial CFG (PCFG) with which feature extraction is performed. The features are used to train a Variational Graph Autoencoder (VGAE) model \cite{kipf2016variational} that encodes CFGs into a lower-dimensional space (graph embedding). The legitimate control-flow path is converted to an embedding which serves as a reference.  Ten legitimate control-flow paths are used to set an appropriate threshold using the directed Hausdorff distance~\cite{hausdorff1914grundzuge} to quantify the difference between two embeddings. During the attestation process, \prover sends a measurement result, a sequence of BBL addresses, executed by the target program to \verifier. \verifier inputs this data to the VGAE model to produce a new embedding. \verifier compares the embedding with the referenced embedding; if their distance exceeds a predefined threshold, then it indicates the detection of attacks. The authors tested their proposal using synthetic ROP and DOP attacks, yielding results of 0.9803\ (ROP) and 0.9101 (DOP) F1-score while maintaining a low False Positive Rate of 0.0319.



\subsection{Whitelist-based Monitoring}

In whitelist-based CFA schemes, \verifier generates an enumeration of all authorised control-flow transfers in an offline phase. In the online phase, \prover sends a sequence of target control-flow transfers taken by the target program to \verifier, represented by (\textit{src}, \textit{dst}) pairs. \verifier matches each (\textit{src}, \textit{dst}) pair against the whitelist; any mismatch indicates the occurrence of a control-flow attack. The approach has been used by a minority of proposals~\cite{RN129,RN169}.

An advantage of whitelist-based measurement is that \verifier has greater visibility over the nature of the control-flow violation, which can help with patching and remediation efforts. Another advantage is that it demands less storage space on \verifier than hash-based approaches.  \verifier still requires a whitelist to store all legitimate control-flow transfers of the target programs, but the path explosion problem is avoided; the storage requirement is proportional to the number of legitimate control-flow transfers.

A concern with whitelist approaches is the need to monitor, and potentially store, control-flow transfers in real-time. This can impose significant time and space requirements on \prover.  Indeed, if the attestation report is sent to \verifier only after the target program terminates, the measurement module must store the sequence in memory until that point. Indeed, a long sequence of control-flow transfers produces a large attestation report size.

One way to mitigate the problem caused by long control-flow transfer sequences is using partial attestation reports. Here, \prover generates a mini-attestation report when specific requirements are met---e.g.\ the configured length of the sequence is reached \cite{LiteHAX}, and particular piece of code finishes \cite{oat}---in order to prevent memory exhaustion. Using partial attestation reports allows \verifier to receive information about control-flow paths once certain portions of the target program have finished execution without waiting for it to terminate. Indeed, partial attestation reports can meet some scenarios with real-time requirements (see ARI~\cite{RN147}). Compression has also been used to minimise the measurement results in attestation reports; for example, ReCFA~\cite{RN129} used Zstandard~\cite{collet2021rfc}, an off-the-shelf compression algorithm, to reduce the size of control-flow transfer sequences.

\subsection{Abstract Execution} \label{sec:Abstract execution-based measurement}
Abstract execution~\cite{RN174} is a method for monitoring authorised program states by simulating or mimicking the execution of the proving program. This is principally performed on the side of \verifier who, after receiving a set of program states from \prover, checks whether it corresponds to a valid set of program states. In the context of CFA, this is similar to whitelist-based measurements. However, rather than representing control-flow transfers as $(src, dst)$,  \verifier simulates the execution of \proverprogram by aggregating control-flow transfers using its CFG. The set of states from \prover is compared with legitimate, reconstructed paths in order to detect the presence of control-flow hijacking attacks. The representation of control-flow transfers more generally depends on different CFA proposals. One way is to represent indirect control-flow transfers by its $dst$ value and conditional branches using a binary encoding, where 0 = not taken and 1 = taken~\cite{LiteHAX}. Another representation involves assigning dedicated IDs to target control-flow transfers~\cite{RN139}.

\begin{figure}
    \centering
    \includegraphics[width=\linewidth]{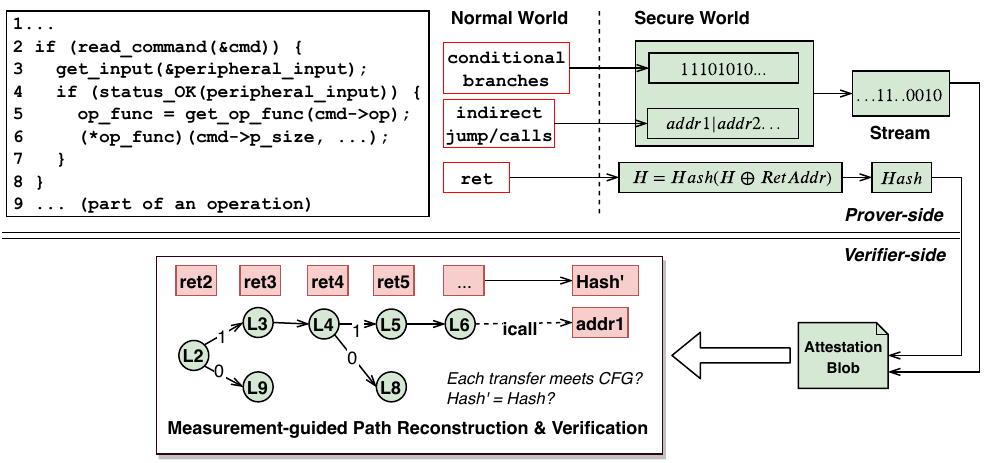}
    \caption{The OAT verification procedure~\cite{oat}.}
    \label{fig:oat-verification-procedure}
\end{figure}

Consider the example of the OAT CFA scheme by Sun et al.~\cite{oat}. Here, \verifier starts abstract execution from the root BBL, or the entry point of \proverprogram's CFG, and follows forward-edge traces. \verifier maintains a simulated call stack to track return addresses in memory, computing hash values as \prover does. Violations are detected when a mismatch is found between the current BBL and its trace, i.e.\ the block ends with an indirect call when the next element is a branch; the computed hashes do not equal known `safe' values; or the addresses are not valid. \verifier terminates the abstract execution when it encounters a BBL whose final instruction is not a control-flow transfer. We illustrate the OAT verification procedure in Figure~\ref{fig:oat-verification-procedure}.

GuaranTEE~\cite{RN139} focuses on intra-platform attestation between two TEE-enabled applications, one which locally verifies the proving application, \verifier' and \prover respectively. We refer the reader back to Figure~\ref{fig:morbitzer} for an overview of the proposal. \verifier' hosts a `control analyser' that collects control-flow transfers passed from \prover over a trampoline. \verifier' has a copy of the known, legitimate CFG that \prover ought to follow when executing the target platform. As control-flow transfers are passed \prover$\to$ \verifier', then \verifier' compares them against the expected transfers from the CFG, recording any differences in an attestation log. Such differences are logged as a hash chain, which tracks ID values corresponding to the endpoints of CFG edges, and encrypted using AES in GCM mode. This data forms part of the attestation report that can be sent to a remote verifier, \verifier.

\subsection{Hybrid Measurement}

We define hybrid measurement as combining more than one of the measurement methods above. Currently, three existing CFA proposals fall into the category, which combine hash-based measurement and abstract execution. LO-FAT~\cite{RN157} uses abstract execution for loop paths and hash-based measurement for the rest of the control-flow path; avoiding hash-based measurements for loops avoids the path explosion problem. OAT~\cite{oat} is the first CFA scheme to explicitly address control-flow and data-only attacks on embedded systems. OAT addresses this through a combination of hash-based measurements and compact execution traces in a TEE (ARM TrustZone).

\begin{figure}
    \centering
    \includegraphics[width=\linewidth]{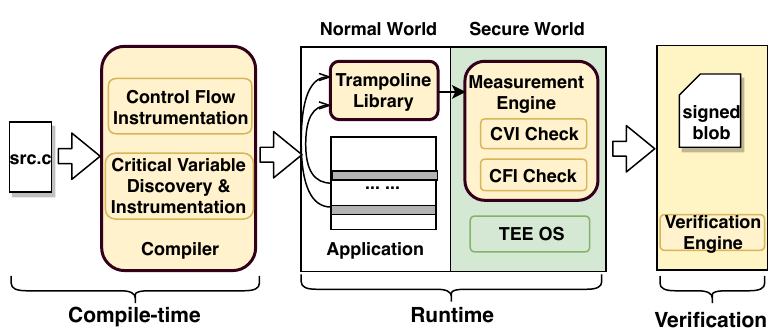}
    \caption{Deployment overview of OAT~\cite{oat}.}
    \label{fig:oat-overview}
\end{figure}

OAT's system overview is given in Figure~\ref{fig:oat-overview}. The source program, \proverprogram, undergoes control-flow and critical variable instrumentation using an LLVM-based compiler.  At run-time, a TEE-based measurement engine consumes application events using a trampoline in the normal world. The engine checks the integrity of `critical variables', which are annotated by developers, using a value-based approach to reduce performance overhead. More specifically, it checks whether the value at an instrumented load instruction is the same as that recorded at the previous store instruction. The engine also measures CFI using a combination of traces for forward-edge control flow transfers (i.e.\ branches and indirect calls/jumps), and hash-based measurements for backward-edge transfers (returns).  Traces and hashes are combined to form an `attestation blob' that is sent to \verifier. After receiving the blob, \verifier abstractly executes \proverprogram using its CFG by starting from its entry BBL to reconstruct the measurement hatch, which we described previously in \S\ref{sec:Abstract execution-based measurement}.

\begin{figure*}
    \centering
    \includegraphics[width=0.93\linewidth]{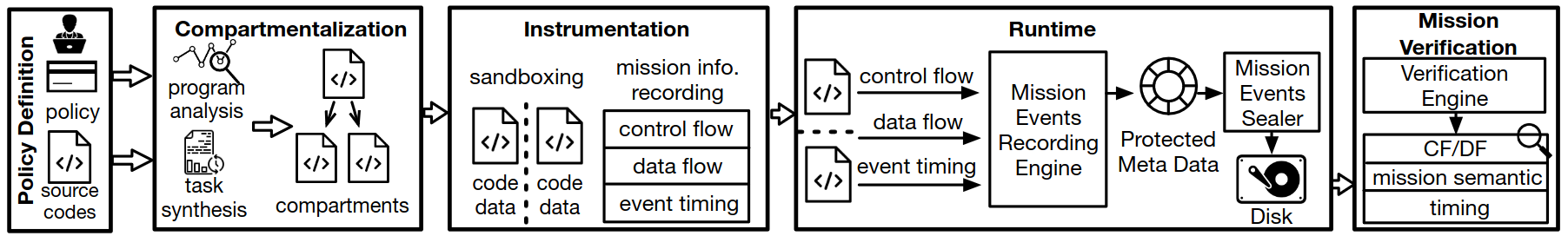}
    \caption{Workflow of ARI by Wang et al.~\cite{RN147}.}
    \label{fig:ari-workflow}
\end{figure*}

\begin{figure}
    \centering
    \includegraphics[width=0.825\linewidth]{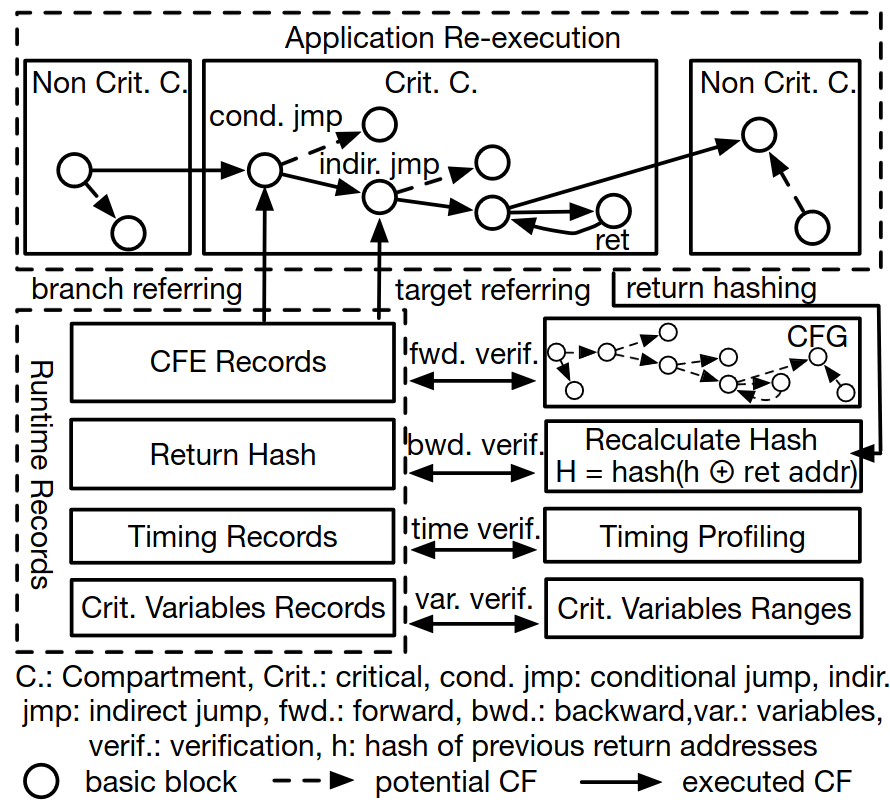}
    \caption{ARI's verification procedure executed by \verifier~\cite{RN147}.}
    \label{fig:ari-verif}
\end{figure}

ARI by Wang et al.~\cite{RN147} focusses on attesting autonomous cyber-physical systems by decomposing mission-critical software into sandboxed `compartments', which are monitored using a combination of control- and data-flow techniques. Compartments are isolated using a memory protection unit or with software-based instrumentation. A trust anchor based on TEEs (ARM TrustZone or RISC-V Keystone) or a dedicated hardware module is suggested. ARI monitors control flows between compartments, e.g.\ hashes of return addresses on the stack; data flows of critical variables; transfers between compartments; and event timestamps. What is defined as a `critical' component or data is set using a policy-based approach. These policies are translated into appropriate instrumentations using an LLVM-based compiler. During \proverprogram's execution, a recording engine captures control and data flow events in the form of an event log, inspired by aircraft flight recorders. The workflow of ARI is given in Figure~\ref{fig:ari-workflow}. In Figure~\ref{fig:ari-verif}, the verification procedure is given, whereby the application is re-executed and different control- and data-flow measurements are calculated and compared with those sent by \prover.

The abstract execution methods of OAT~\cite{oat} and ARI \cite{RN147} use similar techniques. Both proposals use simulated execution on \verifier for verifying forward control-flow transfers, while hash-based measurements assess backward control-flow transfers. The basic steps are as follows: 
\begin{enumerate}
    \item Within \proverprogram, when a control-flow transfer is executed, the measurement module checks the type of transfer and performs the  corresponding operations:
    \begin{enumerate}
        \item For conditional branches, it outputs $\{0,1\}$ to indicate whether or not the branch is taken.
        \item For indirect jumps and function calls, the measurement module outputs the BBL \textit{dst}.
        \item For a return instruction, it records the return address, $ret_{addr}$, and computes a hash value as shown in Equation~\ref{eq:hash-comp}.
    \end{enumerate}	
\begin{equation}
H_{new}=H(H_{prev} \oplus ret_{addr})
\label{eq:hash-comp}
\end{equation}

    \item \prover sends a sequence of control-flow transfers and the final hash value as an attestation report to \verifier.

    \item \verifier simulates the call stack to track return addresses through abstract execution. When the execution encounters a BBL that ends with procedure call, and if the control-flow transfer is authorised, the function's return address is pushed onto the stack. Later, when abstract execution reaches a BBL which ends with a return instruction, \verifier inspects the return address from the top of the stack to determine the direction of execution, before popping the return address from the stack. The hash is calculated in the same way as the measurement module does during runtime.

    \item If no violations are detected during abstract execution, \verifier compares the computed hash value with the one reported by \prover. A match indicates no control-flow attacks were detected, while a mismatch suggests the occurrence of an attack.
\end{enumerate}

We note that the results of abstract execution are used alongside known static values. In ARI, data-flow integrity is assessed simultaneously using a value-based approach (i.e.\ inspecting the value of critical variables between store and load operations). Supplementary information, such as timing profiles, may also be used~\cite{RN147} (see Figure~\ref{fig:ari-verif}).

Using abstract execution for forward control-flow transfers and hash-based measurement for backwards control-flow transfers inherits advantages of both approaches. Firstly, it provide space-efficiency by shortening the length of measurement results. Sun et al.~\cite{oat} argued that target programs typically execute a large number of return instructions, i.e.\ backward control-flow transfers. Hashing is naturally space-efficient over listing them sequentially, significantly reducing the size of the measurement result. Indeed, OAT reported an average size reduction of 97\%~\cite{oat}. 

A second advantage is that it ameliorates a major drawback of hash-based measurements: the requirement for \verifier to store all possible control-flow paths and their corresponding hash values. Return addresses, which serve as input of hash computation, are obtained by abstract execution based on memory addresses of function calls. As such, \verifier does not have to know possible backward control-flow paths during the offline phase, but instead only the final value. 

However, hybrid methods still inherit a significant disadvantage of hash-based approaches. Specifically, if the hash value from the measurement result does not match the one obtained from abstract execution, then \verifier does not know what backward control-flow path was violated. This limits \verifier's ability to precisely identify potential vulnerabilities that exploit return addresses (e.g.\ as used by ROP attacks).
\section{Evaluation} \label{Summary}

A significant number of control-flow attestation proposals have been published in the past decade. To aid in future research, we consolidate and compare these publications, their features and shortcomings against a common set of criteria. In total, we have identified \npapers distinct proposals that fall under the definition of control-flow attestation published between the years 2016 and 2024. In this section, we describe the criteria used to compare such works based broadly on the sections presented hitherto.

\begin{table*}[H]
\caption{A summary of CFA proposals using the comparison criteria from \S\ref{sec:comparison-criteria}.}
\centering
\resizebox{\textwidth}{!}{%
\begin{threeparttable}
\begin{tabular}{ | p{0.1255\textwidth} | p{0.0352\textwidth} | p{0.145\textwidth} | p{0.1\textwidth} | p{0.145\textwidth} | p{0.02\textwidth} | p{0.1\textwidth} | p{0.13\textwidth} | p{0.127\textwidth} | }
 \toprule
  \textbf{Proposal} & \textbf{Year} & \textbf{Trust Anchor} &  \textbf{Measurement Approach} &  \textbf{Instrumentation Method} & \textbf{FI?} & \textbf{\prover-\verifier Paradigm} &  \textbf{Target Control-Flow Transfers} & \textbf{Control-Flow Attacks}  \\ 
  \midrule
  \emph{C-FLAT}~\cite{cflat}$^{\Diamond}$ & 2016 & TEE (ARM TrustZone) & Hash-based & Binary Rewriting & \cmark & Conventional & \ref{lab:all-control-flow-transfers} & \ref{lab:code injection attacks}\allowbreak\ref{lab:code reuse attacks}\allowbreak\partialreq{\ref{attack:decision-making}} \\
  \hline
  \emph{LO-FAT}~\cite{RN157} & 2017 & CPU Extensions & Hybrid & \hwinstrumentation & \cmark & Conventional & \ref{lab:all-control-flow-transfers} & \ref{lab:code injection attacks}(MP)\allowbreak\ref{lab:code reuse attacks}\allowbreak\partialreq{\ref{attack:decision-making}} \\
  \hline
  \emph{ATRIUM}~\cite{RN144} & 2017 & CPU Extensions & Hash-based & \hwinstrumentation & \cmark & Conventional & \ref{lab:all-control-flow-transfers} & \ref{lab:code injection attacks}\allowbreak\uncertain{\ref{lab:code reuse attacks}}\allowbreak\partialreq{\ref{attack:decision-making}} \\
  \hline
  \emph{LiteHAX}~\cite{LiteHAX} & 2018 & CPU Extensions & Abstract Execution & \hwinstrumentation & \cmark & Continuous Reporting & \ref{lab:all-control-flow-transfers} & \ref{lab:code injection attacks}\allowbreak\uncertain{\ref{lab:code reuse attacks}}\allowbreak\ref{attack:decision-making}\allowbreak\ref{attack:dop} \\
  \hline
  \emph{CHASE}~\cite{RN138} & 2019 & CPU Extensions & Hash-based (MSH) & \hwinstrumentation & \cmark & LVRR & \uncertain{\ref{lab:mixed-cf}} & \ref{lab:code injection attacks}\allowbreak\ref{lab:code reuse attacks}\allowbreak\partialreq{\ref{attack:decision-making}} \\
  \hline
  \emph{DIAT}~\cite{diat} & 2019 & TEE* & Hash-based (MSH) & \uncertain{Binary Rewriting}  & \xmark & Collective & \ref{lab:mixed-cf} & \ref{lab:code injection attacks}$\ddagger$\allowbreak\ref{lab:code reuse attacks}$\ddagger$\allowbreak\partialreq{\ref{attack:decision-making}$\ddagger$}\allowbreak\partialreq{\ref{attack:dop}$\ddagger$} \\ 
  \hline
  \emph{G\&R}~\cite{gedenrasmussen} & 2019 & Hardware Module & \hwinstrumentation & \hwinstrumentation & \cmark & LVRR & \ref{lab:conditional-direct-jump}\allowbreak\ref{lab:unconditional-direct-jump}\allowbreak\ref{lab:direct-calls}\allowbreak\ref{lab:return-instructions} & \ref{lab:code injection attacks}(MP)\allowbreak\ref{lab:code reuse attacks}\P\allowbreak\ref{attack:decision-making}\allowbreak\ref{attack:dop} \\
  \hline
  \emph{Liu et al.}~\cite{RN169} & 2019 & TEE (ARM TrustZone) + PUF & Whitelist & Binary Rewriting & \cmark & Conventional & \ref{lab:conditional-direct-jump}\allowbreak\ref{lab:indirect-jumps}\allowbreak\ref{lab:indirect-calls}\allowbreak\ref{lab:return-instructions}  & \ref{lab:code injection attacks}\allowbreak\ref{lab:code reuse attacks}\allowbreak\partialreq{\ref{attack:decision-making}}\\
  \hline
  \emph{MGC-FA}~\cite{RN145} & 2019 & TEE (ARM TrustZone) & Hash-based & Compiler (GCC and Capstone) & \cmark & Conventional & \ref{lab:mixed-cf} & \uncertain{\ref{lab:code injection attacks}$\pm$\allowbreak\ref{lab:code reuse attacks}$\pm$}\allowbreak\partialreq{\ref{attack:decision-making}$\pm$} \\
  \hline
  \emph{RADIS}~\cite{RN156} & 2019 & TEE* & Hash-based & \uncertain{Dynamic}
& \xmark & Collective & \ref{lab:all-control-flow-transfers}  & \ref{lab:code injection attacks}\allowbreak\uncertain{\ref{lab:code reuse attacks}}\\ 
  \hline
  \emph{ScaRR}~\cite{scarr} & 2019 & Kernel & Hash-based & Compiler (LLVM) & \cmark & Conventional & \ref{lab:conditional-direct-jump}\allowbreak\ref{lab:direct-calls}\allowbreak\ref{lab:indirect-jumps}\allowbreak\ref{lab:indirect-calls}\allowbreak\ref{lab:return-instructions}  & \ref{lab:code injection attacks}\allowbreak\ref{lab:code reuse attacks} \\
  \hline
  \emph{CFPA}~\cite{RN176} & 2019 & TEE* & $\bigstar$ & $\bigstar$ & \xmark & Conventional & $\bigstar$ & \ref{lab:code injection attacks}\allowbreak\ref{lab:code reuse attacks}\allowbreak\partialreq{\ref{attack:decision-making}} \\ 
  \hline
  \emph{DO-RA}~\cite{RN132} & 2020 & TEE (ARM TrustZone) & Hash-based & Binary Rewriting & \cmark & Conventional & \ref{lab:conditional-direct-jump}\allowbreak\ref{lab:indirect-jumps}\allowbreak\ref{lab:indirect-calls}\allowbreak\ref{lab:return-instructions}  & \ref{lab:code injection attacks}(MP)\allowbreak\uncertain{\ref{lab:code reuse attacks}}\allowbreak\ref{attack:decision-making} \\
  \hline
  \emph{LAPE}~\cite{RN135} & 2020 & MPU & Hash-based & Compiler (LLVM) & \cmark & Conventional & \ref{lab:unconditional-direct-jump}\allowbreak\ref{lab:indirect-calls}  & \ref{lab:code injection attacks}\allowbreak\ref{lab:code reuse attacks} \\
  \hline
  \emph{OAT}~\cite{oat}$^{\Diamond}$ & 2020 & TEE (ARM TrustZone) & Hybrid & Compiler (LLVM) & \cmark & Conventional & \ref{lab:mixed-cf}  & \ref{lab:code injection attacks}$\ddagger$(MP)\allowbreak\uncertain{\ref{lab:code reuse attacks}$\ddagger$}\ref{attack:decision-making}$\ddagger$\allowbreak\ref{attack:dop}$\ddagger$ \\
  \hline
  \emph{ARCADIS}~\cite{ARCADIS} & 2021 & ROM & Hash-based & \uncertain{Binary Rewriting} & \cmark & Collective  & \ref{lab:all-control-flow-transfers}  & \ref{lab:code injection attacks}\allowbreak\ref{lab:code reuse attacks} \\
  \hline
  \emph{GACFA}~\cite{RN158} & 2021 &TEE (Intel SGX) & Hash-based & Dynamic (Intel Pin) & \cmark & Conventional & \ref{lab:mixed-cf} & \ref{lab:code injection attacks}$\pm$\allowbreak\ref{lab:code reuse attacks}$\pm$\allowbreak\partialreq{\ref{attack:decision-making}$\pm$}\\ 
  \hline
  \emph{ReCFA}~\cite{RN129}$^{\Diamond}$ & 2021 & Kernel & Whitelist & Binary Rewriting & \cmark & Conventional & \ref{lab:conditional-direct-jump}\allowbreak\ref{lab:direct-calls}\P\allowbreak\ref{lab:indirect-jumps}\allowbreak\ref{lab:indirect-calls}\allowbreak\ref{lab:return-instructions} & \ref{lab:code injection attacks}(MP)\allowbreak\ref{lab:code reuse attacks}\\ 
  \hline
  \emph{Tiny-CFA} \cite{RN160} & 2021 & Hardware Module & \hwinstrumentation & Binary Rewriting & \cmark & Conventional & \ref{lab:all-control-flow-transfers} & \ref{lab:code injection attacks}\allowbreak\partialreq{\ref{lab:code reuse attacks}}\allowbreak\ref{attack:decision-making} \\
  \hline
  \emph{Papamartzivanos et al.}~\cite{RN164} & 2021 &Trace Module (Intel PT) & \hwinstrumentation & \hwinstrumentation & \cmark & Conventional & \ref{lab:conditional-direct-jump}$\dag$\ref{lab:direct-calls}\ref{lab:return-instructions}$\dag$ & \ref{lab:code injection attacks}\allowbreak\ref{lab:code reuse attacks} \\
  \hline
  \emph{BDFCFA}~\cite{BDFCFA} & 2022  &TEE (Intel SGX) & Hash-based & Dynamic (Intel Pin) & \cmark & Conventional & \ref{lab:all-control-flow-transfers}  & \ref{lab:code injection attacks}(MP)\allowbreak\partialreq{\ref{lab:code reuse attacks}}\allowbreak\partialreq{\ref{attack:decision-making}} \\
  \hline
  \emph{C-FLAT Linux}~\cite{ben2022nanovised} & 2022 & Kernel & Hash-based & Binary Rewriting & \cmark & Conventional & \ref{lab:all-control-flow-transfers} & \ref{lab:code injection attacks}\allowbreak\ref{lab:code reuse attacks}\allowbreak\partialreq{\ref{attack:decision-making}}\\
  \hline
  \emph{CFRV}~\cite{RN134} & 2022 & TEE (ARM TrustZone) & Hash-based & Compiler (LLVM) & \cmark & Collective & \ref{lab:all-control-flow-transfers} & \ref{lab:code injection attacks}\allowbreak\partialreq{\ref{lab:code reuse attacks}} \\ 
  \hline
  \emph{GuaranTEE}~\cite{RN139}$^{\Diamond}$ & 2022 &TEE (Intel SGX) & Abstract Execution & Compiler (LLVM) & \cmark & LVRR & \ref{lab:direct-calls}\allowbreak\ref{lab:indirect-jumps}\allowbreak\ref{lab:indirect-calls}\allowbreak\ref{lab:return-instructions}  & \ref{lab:code injection attacks}\allowbreak\partialreq{\ref{lab:code reuse attacks}}\\
  \hline
  \emph{ACFA}~\cite{caulfield2023acfa}$^{\Diamond}$ & 2023 & Hardware Module (On-chip)   & Abstract Execution & \hwinstrumentation & \cmark & Continuous Reporting & \ref{lab:all-control-flow-transfers} & \ref{lab:code injection attacks}\allowbreak\ref{lab:code reuse attacks} \\
  \hline
  \emph{ARI}~\cite{RN147}$^{\Diamond}$ & 2023 & TEE* & Hybrid & Compiler (LLVM) & \cmark & Conventional & \ref{lab:mixed-cf}  & \ref{lab:code injection attacks}$\ddagger$\allowbreak\ref{lab:code reuse attacks}$\ddagger$\allowbreak\ref{attack:decision-making}$\ddagger$\\
  \hline
  \emph{BLAST}~\cite{yadav2023whole} & 2023 & TEE (ARM TrustZone) & Path Log & Compiler (LLVM) & \cmark & Conventional & \uncertain{\ref{lab:all-control-flow-transfers}}  & \ref{lab:code injection attacks}(MP)\allowbreak\uncertain{\partialreq{\ref{lab:code reuse attacks}}} \\
  \hline
  \emph{ISC-FLAT}~\cite{RN149}$^{\Diamond}$ & 2023 &TEE (ARM TrustZone) & Path Log & Binary Rewriting & \cmark & Conventional & \ref{lab:all-control-flow-transfers} & \partialreq{\ref{lab:code injection attacks}}\allowbreak\partialreq{\ref{lab:code reuse attacks}}\\
  \hline
  \emph{ZEKRA}~\cite{debes2023zekra}$^{\Diamond}$ & 2023 & Hardware Module* & Hash-based & \hwinstrumentation* & \xmark & VC & \ref{lab:all-control-flow-transfers} & \ref{lab:code injection attacks}\allowbreak\ref{lab:code reuse attacks}\\
  \hline
  \emph{RAGE}~\cite{chilese2024one} & 2024 & TEE* & Path Log & Dynamic (DynamoRIO) & \xmark& Conventional &\ref{lab:all-control-flow-transfers} & \partialreq{\ref{lab:code reuse attacks}$\pm$}\allowbreak\partialreq{\ref{attack:dop}$\pm$} \\\hline
  \emph{LightFAt}~\cite{gonzalez2024lightfat} & 2024 & PMU & Path Log & Binary Rewriting & \cmark & Conventional &\ref{lab:all-control-flow-transfers} & \ref{lab:code injection attacks}$\pm$$\ddagger$\ref{lab:code reuse attacks}$\pm$$\ddagger$ \\
  \bottomrule
\end{tabular}
\begin{tablenotes}
\item *: Recommended for a complete implementation but not evaluated. \hwinstrumentation: Hardware detection circuit. $\dag$: Suspected based on publicly available information. \P: Partially addresses. $\bigstar$: This work sketches a property-based CFA scheme and high-level requirements; it is neither implemented nor evaluated. $\ddagger$: Applies to only portions of the target program. $\pm$: Uses a probabilistic approach. VC: Verifiable computation. $\Diamond$: Open-source implementation available. LVRR: Local Verification with Remote Reporting.
\end{tablenotes}
\end{threeparttable}
}
\label{table:Summary of CFA proposals}
\end{table*}

\subsection{Comparison Criteria}
\label{sec:comparison-criteria}
Table~\ref{table:Summary of CFA proposals} summarises the CFA proposals found in the state of the art in chronological order. We extracted the key features from each proposal based on the areas presented previously in Figure~\ref{fig:taxonomy_scope}. More specifically, we consider the following criteria:

\begin{itemize}
    \item \textbf{Trust anchor}: The root-of-trust upon which the proposal relies in order to uphold its security, e.g.\ a TEE, custom CPU extensions, an external hardware module, or another method described in \S\ref{sec:trust-anchors}. 
    \item \textbf{Measurement approach}: The way in which control flow is measured during the execution of the target program, as described in \S\ref{Measurement}.
    \item \textbf{Instrumentation method}: The method by which the target program is modified to capture control-flow events from \S\ref{sec:instrumentation}.
    \item \textbf{Full implementation?}: We observed that several proposals make recommendations for ideal trust anchors and instrumentation methods, but which are neither implemented nor evaluated in the proposal. We make this distinction explicit in our comparison.
    \item \textbf{\prover-\verifier paradigm}: We give the prover-verifier (\prover-\verifier) paradigm used by the proposal; for example, whether it uses a conventional remote attestation or another approach described in \S\ref{sec:prover-verifier-paradigms}.
    \item \textbf{Target control-flow transfers}: The types of control-flow transfers that are monitored by the proposal using those discussed in \S\ref{sec:control-flow-types}.
    \item \textbf{Target control-flow attacks}: The types of control-flow attacks addressed by the scheme, which were discussed in \S\ref{sec:cfa-attacks}. We further annotate this column as follows: a `MP' label is used where the proposal assumes a memory protection mechanism to be in place, such as data execution prevention or non-writable code regions. $\ddagger$ indicates that the proposal monitors only specific sections of the target program, not the program in its entirety. $\pm$ is used when the proposal uses a machine learning or statistical model to detect control-flow attacks; that is, whether an attack is detected is probabilistic, not a certainty.
\end{itemize}

\subsection{Discussion and Analysis}

Some trends can be observed in the development of CFA proposals. Firstly, it is noteworthy that the use of CPU extensions and hardware modules as trust anchors and instrumentation methods were used frequently in the earlier literature (2017--2019)~\cite{RN157,RN144,LiteHAX,RN138,gedenrasmussen,RN169}. Such proposals require deep modifications to open-source CPU or SoC designs, which are not commercially available. In more recent work, there is a shift towards those that rely on off-the-shelf technologies, such as proprietary trace modules, memory protection units, performance monitoring units, and kernel-mode implementations (2020--2024)~\cite{gonzalez2024lightfat,debes2023zekra,RN164,ben2022nanovised,RN129,RN135}. The use of proprietary, hardware-assisted methods, such as Intel SGX and ARM TrustZone, has remained dominant over the years, representing the greatest proportion of CFA proposals (16/31, $\approx$52\%)~\cite{cflat,diat,RN169,RN145,RN145,RN176,RN132,oat,RN158,BDFCFA,RN134,RN139,RN147,yadav2023whole,RN149,chilese2024one}. 

There is a clear tension with the use of proprietary technologies and custom architectures. The former tends to be widely available and easily accessible to researchers and potential end-users, e.g.\ software developers. However, this comes at the cost of limited flexibility in monitoring control-flow transfers close to hardware. Custom architectures address this by monitoring events close to the CPU's instruction pipeline, providing the greatest degree of granularity, but at the cost of lacking compatibility with consumer off-the-shelf platforms.

Many CFA schemes claimed the ability to detect code injection attacks, i.e.~\ref{lab:code injection attacks}. However, a significant minority of proposals (7/31, $\approx$23\%) do not provide this due to features bestowed by the scheme itself. Rather, they make assumptions regarding the availability of mechanisms that enforce some aspect of the target program's immutability; for example, using executable-space, data execution prevention, and other memory protection mechanisms~\cite{RN157,gedenrasmussen}. LO-FAT, for instance, assumes that \emph{``the adversary cannot modify
program code at run-time (marked as \texttt{rx})''} (p.\ 3,~\cite{RN157}) as an implicit requirement for a memory protection mechanism. Geden \& Rasmussen~\cite{gedenrasmussen} requires non-writable code regions and non-executable data regions. In a TEE-based scheme, OAT~\cite{oat} assumes that the attackers cannot tamper with the instrumented code or the trampoline library in the normal world, citing some memory protection methods for embedded devices as potential solutions~\cite{clements2017protecting,kim2018securing}.\footnote{We note that, in some cases, it is possible to bypass memory protection mechanisms, e.g.\ DEP, using ROP-style attacks. To this end, we refer the reader to work by G\"{o}ktas et al.~\cite{goktas2014out}.}

We observe that ROP attacks are the primary threat in schemes that address code reuse attacks, \ref{lab:code reuse attacks}. Such proposals add detection mechanisms to control-flow transfers that return from functions. Some work, e.g.\ C-FLAT~\cite{cflat}, address a particular variant in which a function is invoked from multiple calling locations. Suppose there are two legitimate control-flow paths, $A \to B \to C$ and $D \to B \to E$, where $B$ is a vulnerable function. An attacker may perform an attack to alter the control-flow path to $D\to B\to C$, which still visits legitimate CFG nodes and may not be detected as an illicit path. This has been addressed through the use of shadow stacks~\cite{gedenrasmussen,RN129,RN169,scarr}. When a call-based control-flow transfer is executed, if its target address is valid, then its call-after point will be pushed on this shadow stack. For a return-based transfer, if its target address is valid based on CFG, then \verifier checks if it is identical to the top element of the shadow stack. If it is not, a control-flow hijacking attack would be detected. If it matches, the top element is popped from the stack. Call-return matching is another solution in which \verifier marks legitimate pairs of call and return control-flow transfers within the scheme's offline phase, which are measured at run-time~\cite{cflat}.

All CFA proposals claimed that code reuse attacks can be detected to some extent; however, only 11/31 ($\approx$35\%) specifically mentioned call-return matching or shadow stacks as well-known countermeasures~\cite{cflat,gedenrasmussen,LiteHAX,RN138, RN129,oat,RN147,RN149,RN169,debes2023zekra,scarr}. In some instances, details about the call-return matching implementation have not been discussed fully~\cite{LiteHAX}. The propensity to focus on ROP attacks is notable. In recent years, other code reuse attacks have been developed, such as COP and JOP attacks, while only ROP attacks have been demonstrated. In future work, we urge that researchers demonstrate a wider range of code reuse attacks, such as COP, JOP, return-to-libc, \emph{and} ROP attacks.

In a related challenge, many proposals attest the program at launch-time, which can give rise to TOCTOU attacks if an adversary conducts an attack before or after an attestation challenge. Continuous reporting proposals explicitly aim to address this challenge~\cite{LiteHAX,caulfield2023acfa} (see \S\ref{Uni-directional Reporting}), where \prover actively sends measurements to \verifier.  However, this paradigm is used by a very small number of proposals (2/31, $\approx$6\%). Conventional attestation schemes may still detect this by \prover storing measurements until they are requested by \verifier (see ARCADIS~\cite{ARCADIS}). Yet, this relies on \verifier determining an appropriate timeframe to issue the challenge. Simply performing the challenge at launch-time, for example, would be insufficient in detecting transient attackers. 

A final point of note is the relative lack of focus on data-oriented programming (DOP) attacks (\ref{attack:dop}). Only 5/31 ($\approx$16\%) proposals address \ref{attack:dop}. We have mentioned that DOP attacks represent a highly sophisticated attacker (see \S\ref{sec:cfa-attacks} and  \cite{LiteHAX}). The principal focus of CFA schemes is on direct \emph{control} flow attacks, \ref{lab:code injection attacks} and \ref{lab:code reuse attacks}, rather than \emph{data}-oriented attacks, even though the latter can influence the former indirectly. Indeed, some schemes explicitly declare data-flow attacks to be out-of-scope, even though their potential effects on control-flow are recognised~\cite{cflat,RN176}. Preventing data-oriented attacks is a burgeoning area of research~\cite{cheng2021exploitation}, and it is our belief that designing attestation schemes that detect control- \emph{and} data-flow attacks is an important direction going forwards.

\section{Open Problems and Recommendations}
\label{sec:open-problems}

This section discusses research gaps and makes suggestions and recommendations for future work in the field.

\subsection{Platform Dependencies}

We observed that many CFA schemes rely on proprietary platform features, such as Intel PT for measuring control-flow traces, and Arm TrustZone for protecting code integrity. This can negatively affect the potential longevity of proposals if certain features are made obsolete by platform vendors. We note, for example, that Intel SGX was recently discontinued for consumer devices (Intel Core CPUs), limiting the applicability of relevant CFA schemes that depend on it. It is possible that analogous technologies may be present on other platforms, e.g.\ Arm TrustZone over Intel SGX, and that the main idea of the scheme can still be applied. Nevertheless, not every TEE offers the same security properties. We urge caution when translating schemes to other platforms and the impact this has on reducing the CFA scheme's security. 

\begin{itemize}
    \item[$\diamond$] \textbf{Recommendation}: We recommend that authors relying on platform-specific security features for their CFA schemes offer alternatives for those features on other platforms. Alternatively, authors ought to consider generic constructions that are agnostic to particular platform features.
\end{itemize}

\subsection{Limited Deployment Scenarios}

During the course of this work, it became evident that CFA proposals tend to be evaluated in laboratory conditions or using fixed datasets, e.g.\ SPEC CPU benchmarks, that do not replicate many real-world scenarios. In particular, we recognised that CFA schemes do not consider long-running applications, like those that control sensors and actuators. Such programs may execute a large number of control-flow transfers during run-time. For schemes that measure control-flow transfers, \verifier might receive large attestation reports; in a path-based scheme, \verifier may find it challenging, if not infeasible, to obtain all legitimate control-flow paths due to the enormous state space.  We observed that very few proposals (e.g.\ OAT \cite{oat}) support customisable code attestation, which could alleviate this problem.

A large number of proposals rely on the submission of program inputs in order to trigger a measurement of traversed CFG nodes. This paradigm may cause issues in applications where the program inputs do not traverse affected nodes, e.g.\ affected by a code injection attack.  For example, consider a device that accepts a significant number of inputs, such as a robotic arm or another cyber-physical system (\prover) that communicates with a remote control service (\verifier). What parameters should \verifier submit to \prover? Is it possible that the supplied parameters could unintentionally miss malicious code within a larger complex program? These questions, among others, have received limited attention in the literature. 

It is also conceivable that \prover relies on sensitive data that \verifier cannot reasonably access, such as personally identifiable information or security-sensitive sensor data.  Unfortunately, we found that authors rarely discuss application- or domain-specific implications, even though it may impact the deployability of a CFA scheme in practice.

\begin{itemize}
    \item[$\diamond$] \textbf{Recommendation}: We advise that authors describe the deployment assumptions and some challenges that may arise therein under realistic scenarios. In conjunction with the recommendation in \S\ref{sec:common-benchmarks}, it may be useful to develop evaluation approaches and datasets that reflect real-world problems in common domains of interest, e.g.\ cyber-physical systems.
\end{itemize}

\subsection{Closed-source Implementations}

We also became aware that most CFA proposals are presented without any publicly available artefacts. Only 8/31 ($\approx$26\%) of CFA proposals make their implementation publicly available. This is particularly crucial when a scheme demands extensive low-level changes, such as custom CPU extensions as FPGA/HDL modules, binary-level modifications, custom kernel modules, and new assembly routines. This can impose a severe cost to students and researchers who may wish to rapidly prototype or compare multiple schemes. We also observe that key implementation aspects are sometimes left undefined, such as the method of instrumentation, and some schemes are not fully implemented for evaluation (6/31, $\approx$20\%). 

Ultimately, this may lead to unfaithful implementations and evaluations, impeding the reproducibility of some proposals. Additionally, a few proposals with closed-source implementations do not describe the types of target control-flow transfers or how $\mathcal{V}$ performs CFI checks, which increases the difficulty for us to identify which specific types of control-flow attacks they can detect.

\begin{itemize}
    \item[$\diamond$] \textbf{Recommendation}: We strongly recommend that authors open-source CFA schemes in the pursuit of open science and reproducibility. Schemes that rely on hardware extensions and low-level software modifications are of particular importance.
\end{itemize}

\subsection{On the Need for Common Benchmarks}
\label{sec:common-benchmarks}
A wide variety of CFA schemes have been proposed in the scientific literature with a myriad of system architectures and adversarial models. We note that evaluations tend to be conducted in an \emph{ad hoc} manner, employing different test programs, metrics, and baselines. Consequently, it is difficult to evaluate CFA schemes in order to compare their relative effectiveness.  Some proposals have evaluated common programs. In particular, the open syringe pump is the most common, which is an embedded program for remotely controlling liquid injections in medical applications. This has been used by seven CFA proposals \cite{cflat,RN157,LiteHAX,oat,RN160,RN147,RN149} that are designed for embedded devices. Using more general benchmarks is another choice in evaluation. SNU Real-Time Benchmark is the most common one evaluated by three proposals \cite{RN145,RN158,BDFCFA}. Some proposals \cite{yadav2023whole,debes2023zekra,chilese2024one} published recently opt for Embench-IOT benchmark suite \cite{Embench}. However, they are not designed for CFA schemes.

Besides performance evaluations, the lack of uniform test vectors for evaluating security is also a notable omission in the literature. CRFV \cite{RN134} uses the RIPE \cite{RIPE} benchmark to evaluate the capability of code injection attacks and ROP attack detection. Some proposals (e.g., C-FLAT \cite{cflat}) involve manual attacks that redirect the target program's control flow in the evaluation. Chilese et al. \cite{chilese2024one} evaluate their proposals through a synthetic attack trace generation approach. They develop generators that modify legitimate control-flow paths to compromised ones under ROP and DOP attacks to simulate the two attacks.

\begin{itemize}
    \item[$\bigstar$] \textbf{Research Question}: To what extent can a benchmarking suite be developed for CFA schemes that evaluates different control-flow attacks in a uniform manner?
\end{itemize}

\subsection{Addressing Physical Attacks}

It is widely common for authors to deem physical attacks as outside the scope of their threat models, implicitly or explicitly. This is acceptable where we can safely assume that adversaries are strictly remote or software-based ones. However, many CFA schemes are designed for embedded systems, which may be deployed in largely unsupervised environments, such as industrial control systems and critical national infrastructure. The value of such devices to adversaries may increase the viability of physical attacks, such as fault injection and side-channel attacks. Indeed, the authors of OAT state: \directquote{We trust code running inside the Secure World (e.g., the measurement engine) and assume that attackers cannot break the TrustZone protection. We also trust our compiler and the trampoline code. We assume that attackers cannot inject code in the Normal World or tamper with the instrumented code or the trampoline library}~\cite{oat}. It is widely known that TEEs are susceptible to hardware attacks~\cite{shepherd2021physical}. We pose this as a significant gap in the literature, and urge the development of CFA schemes that consider such threats.

\begin{itemize}
    \item[$\bigstar$] \textbf{Research Question}: To what extent can CFA schemes be made explicitly resilient to hardware attacks, such as fault injections and side-channel analysis?
\end{itemize}

\subsection{Extended Attestation Paradigms}

The vast majority of CFA schemes follow a traditional remote attestation paradigm using a single proving device and verification authority. We recognise that some different approaches to attestation may be of interest to researchers, which were developed for TPM-based platforms but have not been fully realised in the area of control-flow attestation. In particular, property-based~\cite{sadeghi2004property}, direct anonymous~\cite{brickell2004direct}, and mutual attestation~\cite{shepherd2019remote} have not been adopted as potential attestation models for CFA. We believe that these areas are fruitful avenues for researchers in the field.

\begin{itemize}
    \item[$\bigstar$] \textbf{Research Question}: To what extent can other attestation paradigms, such as property-based (PBA), mutual, and direct anonymous attestation (DAA), be integrated into control-flow attestation?
\end{itemize}
\section{Conclusion} \label{sec:conc}
In this paper, we presented a comprehensive analysis of control-flow attestation as it is currently realised in the state of the art, exploring the concepts, solutions, techniques and assumptions of \npapers proposals in the field. We categorised existing CFA proposals based on their key features using a common set of criteria, including their instrumentation method, target control-flow transfers, attacks addressed, attestation paradigm, and more. Based on this, we presented a series of open challenges and recommendations for steering future research in the field. Through this work, we hope to enhance the reader's understanding of the capabilities and considerations involved in designing and analysing control-flow attestation schemes, and the challenges that lie ahead.

\bibliographystyle{ieeetr}
\bibliography{refs}

\newpage
\bio{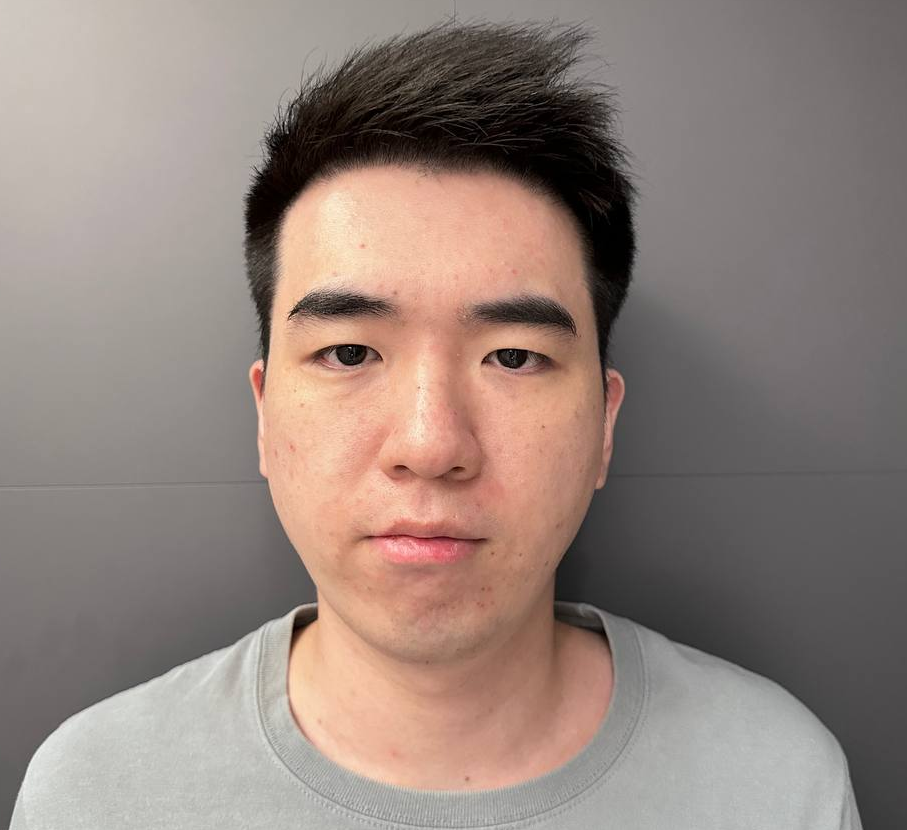}
\textbf{Zhanyu Sha} (B.Sc, M.Sc) is currently a Ph.D.\ student in Information Security at Royal Holloway, University of London. He received his M.Sc.\ in Cyber Security from King's College London. He also earned two B.Sc.\ degrees: one in Information and Computer Science from Xi'an Jiaotong-Liverpool University, China, and another in Computer Science from the University of Liverpool, U.K. His research interests include software security, trusted execution environments, and embedded systems.
\endbio

\bio{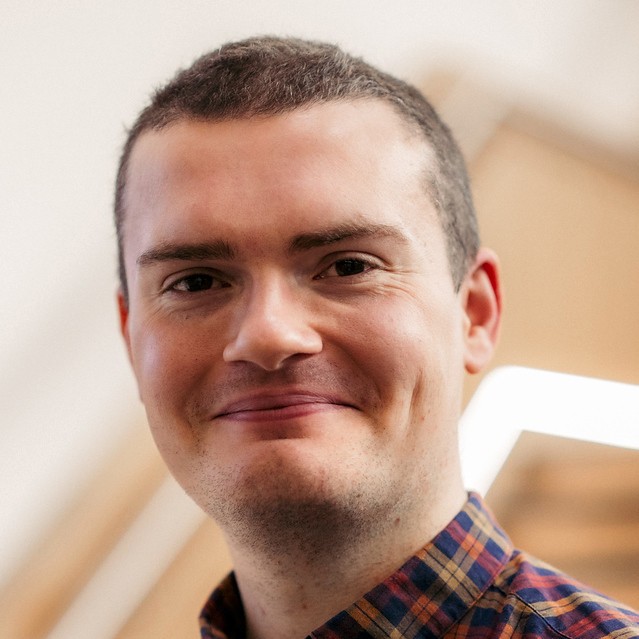}
\textbf{Carlton Shepherd} (B.Sc., Ph.D.) received his Ph.D.\ Information Security from Royal Holloway, University of London, and his B.Sc.\ in Computer Science from Newcastle University. He is currently a Lecturer ($\sim$Assistant Professor) in Computer Science at Newcastle University, and was previously Senior Research Fellow at the Information Security Group at Royal Holloway, University of London. His research interests centre around the security of trusted execution environments (TEEs) and their applications, secure CPU design, embedded systems, applied cryptography, and hardware security.
\endbio%

\bio{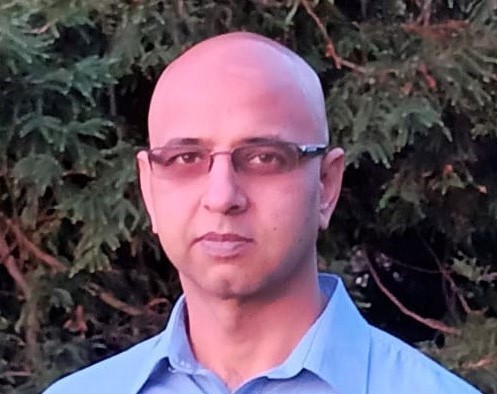}
\textbf{Amir Rafi} (B.Sc., M.Sc.) obtained his M.Sc.\ in Information Security from Royal Holloway, University of London, his B.Sc.\ in Computing from Queen Mary, University of London, and he is currently completing his Ph.D.\ in Information Security at Royal Holloway, University of London. He is a member of the Information Security Group Smart Card and IoT Security Centre (SCC) and was previously Research Assistant at the Information Security Group at Royal Holloway, University of London. His research interests include digital rights management, trusted execution environments and embedded systems security.
\endbio

\bio{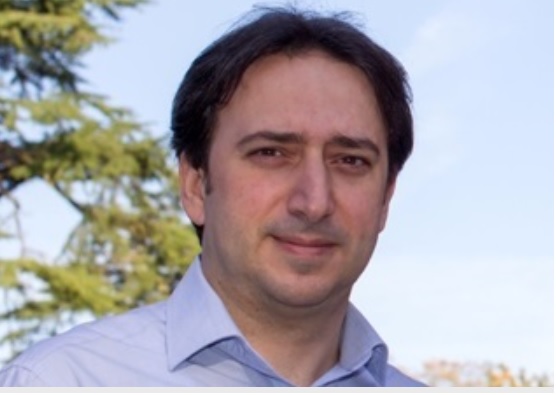}
\textbf{Konstantinos Markantonakis} (B.Sc., M.Sc., MBA, Ph.D.) is a Professor of Information Security in Royal Holloway University of London, and the Director of the Information Security Group Smart Card and IoT Security Centre (SCC). He obtained his B.Sc.\ (Lancaster University), M.Sc., Ph.D.\ (London) and his MBA in International Management from Royal Holloway, University of London. His research interests include smart card security and applications, secure cryptographic protocol design, embedded systems security, autonomous systems and trusted execution environments.
\endbio

\end{document}